\documentclass[twocolumn,floatfix,showpacs,prd,aps,tightenlines,letterpaper,superscriptaddress]{revtex4-1}
\usepackage{amsmath}
\usepackage{graphicx}
\usepackage{psfrag}
\usepackage[usenames]{color}

\usepackage{dcolumn}
\usepackage{bm}
\usepackage{longtable}
\usepackage[mathscr]{eucal}
\usepackage{mathrsfs}
\usepackage{xspace}

\def\msun{\rm M_{\odot}}
\def\kms{\rm km \, s^{-1}}
\def\etal{{\it et al.}}
\def\simlt{\mathrel{\rlap{\lower 3pt\hbox{$\sim$}}\raise 2.0pt\hbox{$<$}}}
\def\simgt{\mathrel{\rlap{\lower 3pt\hbox{$\sim$}} \raise 2.0pt\hbox{$>$}}}
\def\lsim{\mathrel{\rlap{\lower 3pt\hbox{$\sim$}}\raise 2.0pt\hbox{$<$}}}
\def\gsim{\mathrel{\rlap{\lower 3pt\hbox{$\sim$}} \raise 2.0pt\hbox{$>$}}}

\def\msunpc3{\msun~{\rm {pc^{-3}}}}
\newcommand{\be}{\begin{equation}}
\newcommand{\ee}{\end{equation}}
\def\kms{{\rm\,km\,s^{-1}}}

\newcommand{\bea}{\begin{eqnarray}}
\newcommand{\eea}{\end{eqnarray}}
\newcommand{\beq}{\begin{equation}}
\newcommand{\eeq}{\end{equation}}

\usepackage{hyperref}
\hypersetup{
    pdftitle={Exploring the transition from inspiral to merger in black-hole
binary spacetimes},
    colorlinks=false,            
}

\begin{document}

\def\fun#1#2{\lower3.6pt\vbox{\baselineskip0pt\lineskip.9pt
  \ialign{$\mathsurround=0pt#1\hfil##\hfil$\crcr#2\crcr\sim\crcr}}}
\def\lap{\mathrel{\mathpalette\fun <}}
\def\gap{\mathrel{\mathpalette\fun >}}
\def\kms{{\rm km\ s}^{-1}}
\def\vk{V_{\rm recoil}}

\title{Inspiraling black-hole binary spacetimes: Challenges in transitioning from
analytical to numerical techniques}

\author{Yosef~Zlochower}
\affiliation{Center for Computational Relativity and Gravitation, 
Rochester Institute of Technology, Rochester, NY 14623, USA}

\author{Hiroyuki~Nakano}
\affiliation{Department of Physics, Kyoto University, Kyoto 606-8502, Japan}

\author{Bruno~C.~Mundim}
\affiliation{Institut f\"ur Theoretische Physik,
Johann Wolfgang Goethe-Universit\"at, Max-von-Laue-Str. 1,
60438 Frankfurt am Main, Germany}

\author{Manuela~Campanelli}
\affiliation{Center for Computational Relativity and Gravitation, 
Rochester Institute of Technology, Rochester, NY 14623, USA}

\author{Scott~Noble}
\affiliation{Department of Physics and Engineering Physics, The University of Tulsa, Tulsa, OK 74104}

\author{Miguel~Zilh\~ao}
\affiliation{Departament de F\'{\i}sica Fonamental \& Institut de Ci\`{e}ncies del Cosmos, 
  Universitat de Barcelona, Mart\'{\i} i Franqu\`{e}s 1, E-08028 Barcelona, Spain}

\begin{abstract}

We explore how a recently developed  analytical black-hole binary
spacetime can be extended using numerical simulations to go beyond the
slow-inspiral phase. The analytic spacetime solves the Einstein field
equations approximately, with the approximation error becoming progressively
smaller the more separated the binary. To continue the spacetime
beyond the slow-inspiral phase, we need to transition. Such a transition was
previously explored at smaller separations. Here, we perform this
transition at a separation of $D=20M$ (large enough that the
analytical metric is expected to be accurate), and evolve for six orbits.
We find that small constraint violations can have large
dynamical effects, but these can be removed by using a
constraint-damping system like the conformal covariant formulation of
the Z4 system. We find agreement between the subsequent numerical
spacetime and the predictions of post-Newtonian theory for the
waveform and inspiral rate that is within the post-Newtonian
truncation error.

\end{abstract}

 \maketitle

\section{Introduction}\label{sec:Introduction}

The field of numerical relativity (NR) has progressed at a remarkable
rate since the breakthroughs of 2005~\cite{Pretorius:2005gq,
Campanelli:2005dd, Baker:2005vv}, when it first became possible to
simulate the late inspiral, plunge, merger, and ringdown of black-hole
binaries (BHBs). Recently, Lousto and Healy~\cite{Lousto:2014ida}
completed a long-term 50-orbit precessing BHB
simulation using the moving punctures approach, and
Szilagyi {\it et al.}~\cite{Szilagyi:2015rwa}
completed the longest BHB simulation to date: the last
176 orbits for a nonspinning, intermediate-mass-ratio
($m_1/m_2=1/7$) BHB
using the generalized harmonic
approach.
This is a remarkable achievement, 
but the scaling of the inspiral time with the initial separation $T\sim D^4$ 
means that evolving a binary through the long inspiral is prohibitively 
expensive, even for highly efficient codes. Such a simulation becomes even more
expensive when one is interested in performing long-term dynamical evolutions
of the relativistic magnetohydrodynamics (MHD) of circumbinary disks around
inspiraling supermassive BHBs (SMBHBs). This is because the circumbinary gas 
can exhibit significant secular variations on the time scale of 
hundreds to thousands of binary orbits.

In order to make these long-term simulations possible, our group
developed a complementary approach to treat dynamical BHB spacetimes.
In a series of papers~\cite{Noble:2012xz, Gallouin:2012kb,
Mundim:2013vca, Zilhao:2014ida, Zilhao:2013dta}, we used an analytic
spacetime that is an approximate solution to the Einstein field
equations in the inspiral regime to describe the evolution of the
accretion disks surrounding the binary and each of the individual BHs. 

Our initial approach~\cite{Noble:2012xz}, was one in which
relativistic effects were present but relatively small. In the
situation when gravity is weak [$r_g/r = GM/(rc^2) \ll 1$] and motions
are slow [$(v/c)^2 \ll 1$], the post-Newtonian (PN) approximation
gives a very good description of spacetime.  One can then simply
construct a PN metric which takes energy loss from the binary into
account, accurately modeling both the mass loss and inspiral of the
binary~\cite{Blanchet:2013haa}. Using a spacetime accurate to 2.5PN
order [i.e., including terms up to $\sim (r_g/r)^{5/2}$], but
describing the binary orbital evolution to 3.5PN, we demonstrated
that circumbinary disks can track the inspiral of a SMBHB until the
binary practically reaches the relativistic merger regime~\cite{Noble:2012xz}.
The shortcoming of this approach was that the
PN metric was not valid very close to the BHs, and consequently, we
excised any material that fell within 1.5 binary separations. This
prevented us from studying the dynamics of the gas all the way down
through the horizons of each BH.

In a more recent paper~\cite{Mundim:2013vca}, we extended the metric
to cover the full BHB spacetime up to the rapid plunge state.
 We did this by extending the framework
established in Refs.~\cite{Yunes:2005nn, Yunes:2006iw,
JohnsonMcDaniel:2009dq} for constructing a spacetime metric valid
for initial data, i.e., a
metric accurate for all spatial points but in a very small time
interval, to develop a metric valid for arbitrary times.
In this approach, the {\it near zone} (NZ),
i.e., a zone well outside the two BHs, but less than a gravitational
wavelength from the binary, is still described using a PN
expansion. In the {\it far zone} (FZ), i.e., farther than one wavelength
from the binary, the metric is described by a post-Minkowskian (PM)
expansion. Finally, near each BH, i.e., in the  {\it inner
zone} (IZ), the metric
is described using a perturbed Kerr (here Schwarzschild) BH.
The metrics covering the different
zones are smoothly stitched together using asymptotic expansions and transition
functions in their overlapping regions of validity.

This approach allows us to follow inspiraling SMBHBs over hundreds to
thousands of binary orbits, the timescale on which gas accumulates,
without having to solve the Einstein equations numerically.
The numerical advantage here is that the numerical time step is limited
by fluid characteristic speeds, rather than the much faster speed of
light (this advantage is diminished if we want to evolve gas right near
the horizons). 

Here, we explore the possibility of using a hybrid approach; i.e.,
use the analytic metric for the long inspiral down to separations
where our global spacetime metric is still valid, and then transition
to a full numerical simulation using the analytic spacetime as
initial data
(the use of PN techniques to
generate initial data for BHBs was first developed in
Refs.~\cite{Tichy:2002ec, Kelly:2007uc, Kelly:2009js,
JohnsonMcDaniel:2009dq, Mundim:2010hu}). 
The way to do this is to convert our approximate
spacetime prescription into suitable initial data for $3+1$
NR evolutions, evolve the data forward in time, and compare
the orbital evolution, test particle trajectories,
and gravitational radiation output with our
approximate solution. 

Perhaps more well known is the complimentary approach of combining PN
and other analytical
techniques with numerical waveforms to generate highly accurate hybrid
waveforms. Many authors have explored this and, we refer the reader
to Refs.~\cite{Aasi:2014tra, Ajith:2012az, Boyle:2008ge, Buonanno:2007pf,
  Buonanno:2009qa, Campanelli:2008nk, Nakano:2011pb, Pan:2009wj,
  Pan:2011gk, Pan:2013rra, Szilagyi:2015rwa, Taracchini:2012ig,
Taracchini:2013rva} and references therein.

The main motivation of this paper is to develop techniques to smoothly
transition from an analytically evolved spacetime to a numerically
evolved one. By smooth, here we mean that all families of geodesics
passing through the transition region will have continuous second
derivatives. Here we are considering test particle trajectories as
stand-ins for fluid trajectories. In particular, if there is a jump in
the second derivative of the fluid, we can expect a quasiequilibrium
fluid configuration to shock and therefore require a reequilibration
that may take longer than the inspiral time. Of course, bulk binary
dynamics are important too. Therefore, we want the physics of the
inspiral (rate, orbital frequency) to be as unaffected by the
transition as possible.

We note that the use of PN techniques to generate consistent initial
data (i.e., data with the {\it correct} radiation content) provides
the final ingredient proposed in the ``Lazarus approach'' to
generating waveforms~\cite{Baker:2001sf, Campanelli:2005ia}. The
proposal there was to transition from PN to NR techniques and then
from NR to perturbative techniques.

When using the global analytic metric as initial data, the resulting
initial data are
essentially equivalent (there is only a small difference in the
NZ/FZ transition function and the two metric prescriptions coincide at 
$t=0$) 
to the initial data proposed in
Johnson-McDaniel \etal~\cite{JohnsonMcDaniel:2009dq} and first evolved
in Reifenberger and Tichy~\cite{Reifenberger:2012yg}.
Reifenberger and Tichy compared evolutions of
Bowen-York data~\cite{Bowen:1980yu} to several different
analytic initial data constructions, including Johnson-McDaniel \etal.
Our work here extends upon the work of Reifenberger and
Tichy in several ways. (i) We use the full 4-dimensional metric
of~\cite{Mundim:2013vca} to compare the dynamics of the numerically
evolved metric with the analytic one, (ii)
we evolve binaries with separations large enough that the PN metric
and binary dynamics are expected to be accurate, 
and (iii) we find techniques to ameliorate the inaccuracies
associated with evolving these data that were
discovered by Reifenberger and Tichy.
These inaccuracies arise both from  constraint violations,
due to the fact that our global metric solves the Einstein field
equations only approximately,  and from inaccuracy in the PN orbital angular
momentum and inspiral rate, which can lead to eccentricity in the
numerical
binary evolution.

For the current work, we start the full numerical simulations when the
binary is separated by $D=20M$ and evolve for six orbits. As shown in
Ref.~\cite{Lousto:2013oza}, where the authors there explored numerical
simulations of Bowen-York data at separations ranging from $D=100M$ to
$D=20M$, there is good agreement between the predictions of PN theory
and numerical simulations at $D=20M$. Additionally,
simulations starting at $D=20M$ down to merger are possible with our current
codes, as demonstrated in Ref.~\cite{Lousto:2014ida} (such a
simulation would require approximately 1 $10^6$ CPU hours on an
AMD Opteron machine).

This paper is organized as follows: In Sec.~\ref{sec:ID}, we review
how the analytic BHB inspiral metric is constructed, as well as how it
is used to generate $3+1$ initial data. In Sec.~\ref{sec:techniques},
we describe the techniques we used to numerically evolve the spacetime
metric. In Sec.~\ref{sec:simulations}, we provide details on how the
simulations were performed and the key outcomes of these simulations.
In Sec.~\ref{sec:cmp_to_pn}, we compare the results from the numerical
simulation at separations $\sim20M$ with the predictions of
PN theory. Finally, in Sec.~\ref{sec:discussion} we
discuss our results both in terms of the accuracy of the binary
dynamics (e.g., inspiral rate and orbital frequency)
and in
terms of gravitational waveform generation.

Throughout this paper, we use the geometric unit system, where $G=c=1$,
with the useful conversion factor $1 M_{\odot} = 1.477 \; {\rm{km}} =
4.926 \times 10^{-6} \; {\rm{s}}$.

\section{analytic BHB inspiral metric}\label{sec:ID}

In this paper, we restrict our analysis to non-spinning BHs in
quasicircular orbits. In this context, it is useful to provide a
brief review here of our approximate solution to the Einstein field
equations of a BHB spacetime in the inspiral regime~\cite{Mundim:2013vca}.
The inclusion of spins, both
aligned~\cite{Gallouin:2012kb}
and unaligned
in this spacetime framework
will be the subject of future studies.
\begin{figure}[!ht] \begin{center}
\includegraphics[width=.98\columnwidth,clip=true]{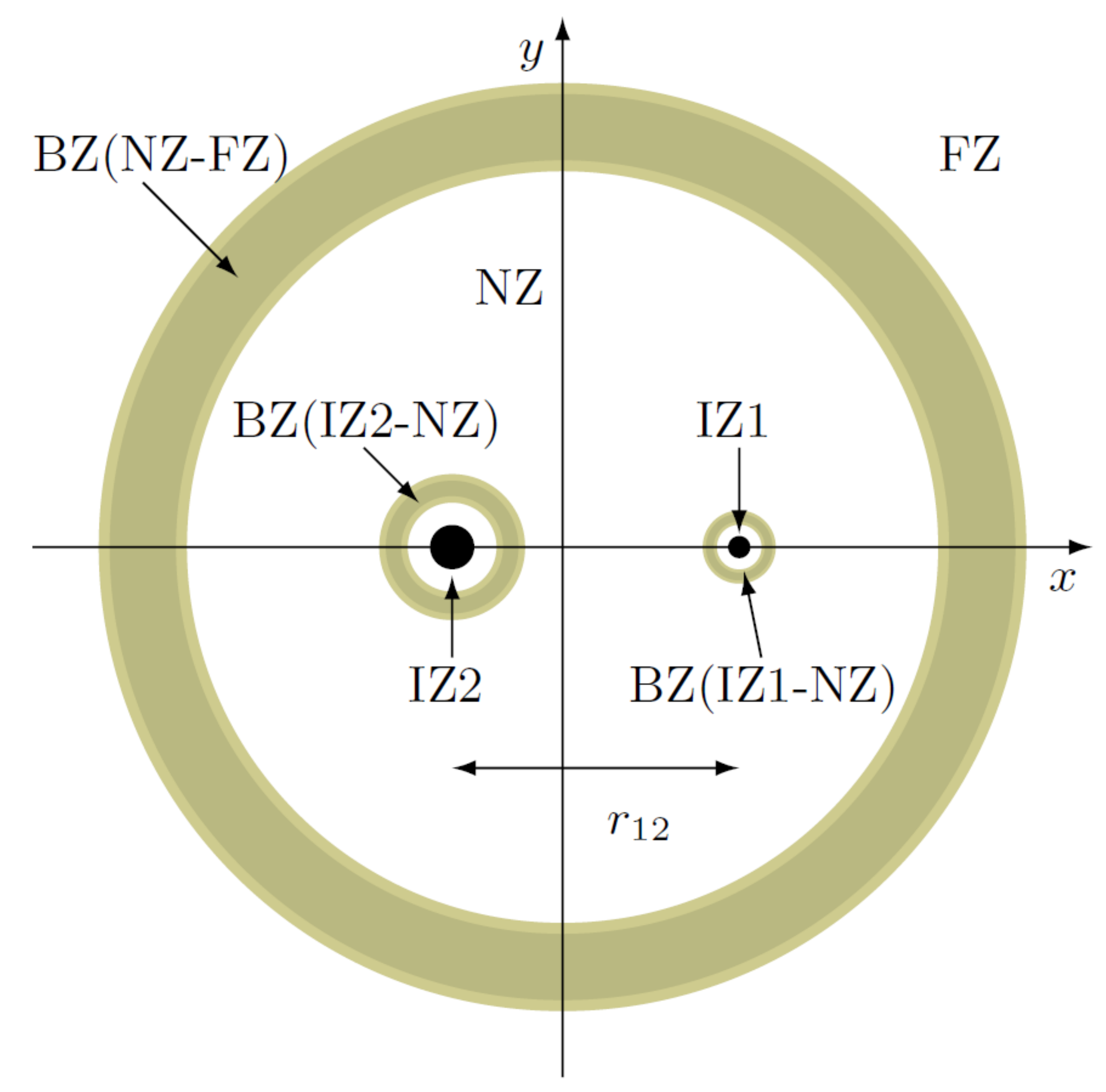}
\end{center}
\caption{Schematic diagram of the zones.
BH1 and BH2 are denoted by solid black dots, where the
orbital separation is $r_{12}$. The BZs are denoted with gray shells,
the outer one representing the FZ/NZ BZ and the two inner ones
representing the NZ/IZ BZs (see also Table~\ref{tab:zones}).
The IZ, NZ and FZ are also shown in the figure.}\label{fig:zones}
\end{figure}
\begin{table}[!ht]
  \caption{Location of the inner, near, and far zones, as well as the
buffer zones joining them. Here $r_{\rm in}$ and $r_{\rm out}$ are
the approximate inner and outer boundaries of a given zone, $m_i$ is
the mass of BH $i$, $r_i$ is the coordinate distance to BH $i$,
$r_{12}$ is the binary separation, and $\lambda$ is the
wavelength of gravitational radiation emitted by the binary.
}
  \label{tab:zones}
\begin{ruledtabular}
  \begin{tabular}{l|ccc}
  Zone           & $r_{\rm in}$ & $r_{\rm out}$ & Region \\
  \hline
  IZ BH1 ($r_1$)    & $0$          & $\ll r_{12}$  & $\cdot$\\
  IZ BH2 ($r_2$)    & $0$          & $\ll r_{12}$  & $\cdot$\\
  NZ ($r_A$)         & $\gg m_A$    & $\ll \lambda$ & $\cdot$\\
  FZ ($r$)            & $\gg r_{12}$ & $\infty$ & $\cdot$ \\
  IZ-NZ BZ & $\cdot$ & $\cdot$ & $m_A \ll r_A \ll r_{12}$ \\
  NZ-FZ BZ & $\cdot$ & $\cdot$ & $r_{12} \ll r \ll \lambda$\\
  \end{tabular}
\end{ruledtabular}
\end{table}

This framework was first introduced in Refs.~\cite{Yunes:2005nn,
Yunes:2006iw, JohnsonMcDaniel:2009dq} as initial data for BHB
evolutions, and was generalized in Ref.~\cite{Mundim:2013vca} to be a
full BHB spacetime.  In this framework, the spacetime is constructed
by asymptotically matching metrics in three different zones
characterizing three different spacetime regions of validity for
different analytic metrics: (i) a far zone (FZ) where the spacetime
can be described by a two-body perturbed flat spacetime with outgoing
gravitational radiation and where retardation effects are fully
accounted for; (ii) a near zone (NZ) which is less than one GW length
from the center of mass of the binary (but not too close to each BH)
that is described by a PN metric (this includes retardation effects at
a perturbative level and binding interactions between the two BHs);
and (iii) inner zones (IZs) that are described by perturbed
Schwarzschild (or Kerr) BHs.  The full spacetime is then constructed
by smoothly transitioning from zone to zone in the so-called buffer
zones (BZs). A schematic diagram of these zones and a table
describing where the zone boundaries are located are provided in
Fig.~\ref{fig:zones} and Table~\ref{tab:zones} (these
were also presented in Refs.~\cite{Gallouin:2012kb, Mundim:2013vca}).

In the sections below, we will refer to these initial data as the {\it
second-order analytical metric}.  It is constructed by asymptotically
matching a 2.5PN metric in the NZ [the matching is only done for
terms up through ${\cal O}(v/c)^4$] to a Schwarzschild
metric with quadrupole (and its time derivatives) and octupole tidal
deformations in the IZ.  As explained in detail in
Ref.~\cite{Mundim:2013vca}, this matching is approximate in the sense
that it does not lead to a formal second-order asymptotic matching in
all metric components for all times.  However, as demonstrated there,
it does lead to a significant improvement against a lower-order
analytic metric, the {\it first-order analytical metric}, which is
constructed by asymptotically matching a 1PN NZ metric [only terms
of order  ${\cal O}(v/c)^2$ are matched] into a Schwarzschild metric
with quadrupole tidal deformations.  The matching for this first-order
metric is exact.  The metric in the FZ is constructed from the
PM expansion over a flat spacetime with source
multipolar decomposition, where the source multipoles are expanded in
the PN approximation up to 2.5PN.  Note that in the PM
formalism, the PN metric in the NZ and the multipolar metric in the FZ
are formally asymptotically matched up to 2.5PN in the NZ-FZ BZ.
The precise orders used for the calculation of the metric
pieces composing these analytical metrics are given in
Table~\ref{tab:orders}.
\begin{table}
   \caption{The orders of the various approximations used in the
     global metric. The virial theorem implies that $m/r$ is taken 
     to be ${\cal O}(v^2/c^2)$,
     and the ${\cal O}$ symbol denotes the highest-order term {\it
   included} in the expansion. Here $g_{ii}$ and $g_{tt}$ refer to the
   diagonal components of the metric; the rest are the off-diagonal
   components. 
Note that the version of the {\it first-order} metric used
in Refs.~\cite{Mundim:2013vca,Zilhao:2014ida} differed from this table in that $g_{tt}$ was
only ${\cal O}(v/c)^2$ there; here, we use resummation techniques
on both the first- and second-order metrics (in Refs.~\cite{Mundim:2013vca,Zilhao:2014ida}
resummation was only used for the {\it second-order} metric).  } \label{tab:orders}
  \begin{tabular}{c|c|c}
    \hline\hline
    & {First order} & {Second order}\\
    \hline
    IZ multipole & $\ell=2$ static & $\ell=2$, $\ell=3$ static\\
    \hline
    NZ $g_{tt}$ & ${\cal O}(v^4/c^4)$ & ${\cal O}(v^7/c^7)$ \\
    NZ $g_{ti}$ & ${\cal O}(v^3/c^3)$ & ${\cal O}(v^6/c^6)$\\
    NZ $g_{ii}$ & ${\cal O}(v^2/c^2)$ & ${\cal O}(v^5/c^5)$\\
    NZ $g_{ij}$ & 0   & ${\cal O}(v^5/c^5)$\\
    \hline
    FZ $g_{tt}$ & ${\cal O}(v^4/c^4)$ & ${\cal O}(v^8/c^8)$ \\
    FZ $g_{ti}$ & ${\cal O}(v^5/c^5)$ & ${\cal O}(v^7/c^7)$\\
    FZ $g_{ii}$ & ${\cal O}(v^3/c^3)$ & ${\cal O}(v^6/c^6)$\\
    FZ $g_{ij}$ & ${\cal O}(v^3/c^3)$ & ${\cal O}(v^6/c^6)$\\
    \hline\hline
  \end{tabular}
\end{table}

\section{Techniques}\label{sec:techniques}

We evolved the BHB initial data using the  {\sc
LazEv}~\cite{Zlochower:2005bj} implementation of the moving puncture
approach~\cite{Campanelli:2005dd, Baker:2005vv} with the conformal
function $W=\sqrt{\chi}=\exp(-2\phi)$ suggested by
Ref.~\cite{Marronetti:2007wz} and the Z4~\cite{Bona:2003fj,
Bernuzzi:2009ex, Alic:2011gg} and BSSN~\cite{Nakamura87, Shibata95,
Baumgarte99} evolution systems. 
Here we use the conformal covariant Z4 (CCZ4) implementation of Ref.~\cite{Alic:2011gg}.
Note that the same technique has been recently applied to the evolution
of binary neutron stars~\cite{Kastaun:2013mv, Alic:2013xsa}.
For the CCZ4 system, we again
used the conformal factor $W$. We used centered eighth-order finite
differencing for all spatial derivatives, a fourth-order Runge-Kutta
time integrator, and both fifth- and seventh-order Kreiss-Oliger
dissipation~\cite{Kreiss73}.

Our code uses the {\sc EinsteinToolkit}~\cite{Loffler:2011ay,
Moesta:2013dna,
einsteintoolkit} / {\sc Cactus}~\cite{cactus_web} / {\sc
Carpet}~\cite{Schnetter-etal-03b, carpet_web}
 infrastructure.  The {\sc Carpet}
mesh refinement driver provides a ``moving boxes'' style of mesh
refinement. In this approach, refined grids of fixed size are arranged
about the coordinate centers of both holes.  The {\sc Carpet} code
then moves these fine grids about the computational domain by
following the trajectories of the two BHs.

We use {\sc AHFinderDirect}~\cite{Thornburg2003:AH-finding} to locate
apparent horizons. We also use the Antenna
code~\cite{Campanelli:2005ia} to calculate the Weyl scalar $\psi_4$. 

We measure the distance between the two BHs using the {\it simple
proper distance} or SPD. The SPD is the proper distance, on a given
spatial slice, 
between the
two BH apparent horizons as measured  along the coordinate line joining the two centers. As
such, it is gauge dependent, but still gives reasonable results
(see Ref.~\cite{Lousto:2013oza} for more details).

\begin{figure}
    \includegraphics[width=.98\columnwidth]{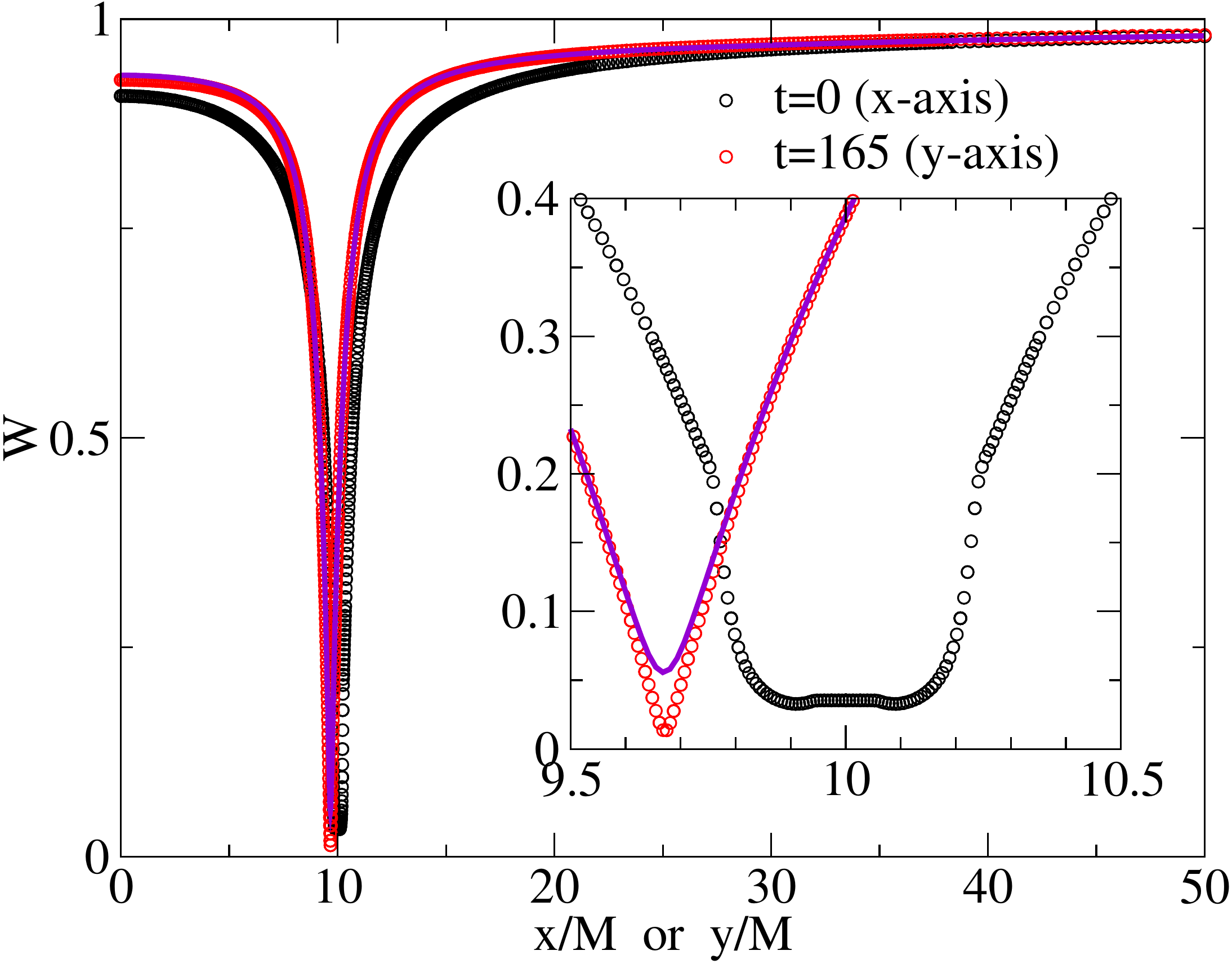}
    \caption{The metric function $W$ at $t=0$ plotted versus $x$
       showing the
      effects of {\it stuffing} the BH interior and the same function
    at $t=165M$ (but plotted versus $y$). The more regular shape of
  $W$ near the center of the BH at $t=165M$ is typical of
  moving puncture simulations (note that the puncture is offset from
  the $y$ axis by $0.01M$).
For reference, the function $W$ for a
Bowen-York puncture simulation (solid curve) 
when the puncture crosses the $x$ axis for the
second time is included (the plot of the Bowen-York data has been
shifted). Note how the {\it stuffed} $W$ appears to evolve to be
very similar to the standard trumpet $W$ typical of puncture initial data.
} \label{fig:stuff}
\end{figure}

To obtain initial data, we use eighth-order finite differencing of the
analytic global metric to obtain the 4-metric and all its first derivatives at
every point on our simulation grid. The finite differencing of the
global metric is constructed so that the truncation error is
negligible compared to the subsequent truncation errors in the full
numerical simulation (here we used finite difference step size of
$10^{-4}$, which is 90 times smaller than our smallest grid size in any
of the numerical simulations discussed below). We then reconstruct the
spatial 3-metric $\gamma_{ij}$ and extrinsic curvature $K_{ij}$ from
the global metric data. Note that with the exception of the
calculation of the extrinsic curvature, we do not use the global 
metric's lapse and shift.  In order to evolve these data, we need to
remove the singularity at the two BH centers. Unlike in the puncture
formalism~\cite{Brandt97b}, the singularities here are true curvature
singularities. We {\it stuff}~\cite{Etienne:2007hr, Brown:2007pg,
Brown:2008sb} the BH interiors in order to remove the singularity.
Our procedure is to replace the singular metric well inside the
horizons with nonsingular (but constraint violating) data through the
transformations 
\begin{eqnarray}
\gamma_{ij} \to f(r)\ \gamma_{ij} \quad i\neq j \,,
\\ 
\gamma_{ii} \to f(r)\ \gamma_{ii} + (1-f(r)) \Xi \,,
\\
K_{ij} \to f(r)\ K_{ij} \,, 
\end{eqnarray}
where
\begin{equation} 
f(r) =
 \left\{\begin{array}{lr}
  0 & r < r_{\rm min} \\
  1 & r > r_{\rm max} \\
  P(r) & r_{\rm min} \leq r \leq r_{\rm max}
 \end{array} \right.,
\end{equation}
$r$ is the distance to a BH center, and $P(r)$ is
a fifth-order polynomial that obeys $P(r_{\rm min}) = P'(r_{\rm
min})=P''(r_{\rm min})=0$, $P(r_{\rm max}) = 1$, $P'(r_{\rm max})
=P''(r_{\rm max})=0$, and $\Xi$ is a large number.  The resulting
data are therefore $C^2$ globally. The parameters $r_{\rm min}$,
$r_{\rm max}$, and $\Xi$ are chosen such that both transitions occur
inside the BHs and so that $W$ varies smoothly with negligible
shoulders in the transition region and is small at the centers.
In Fig.~\ref{fig:stuff} we show the profile of the conformal factor  $W$ at
$t=0M$ and $t=165M$. The former clearly shows the effects of stuffing
while the latter shows that the system appears to evolve to the 
standard moving puncture gauge (i.e., the conformal function $W$
takes on the usual profile for a trumpet slicing just like it does
when using puncture initial data).

\section{Simulations}\label{sec:simulations}

The initial data
parameters for our BHB simulations are given in Table~\ref{table:ID}.
To evolve the {\it second-order analytical} data, we used the following
grid structure:
The coarsest grid spanned $0\leq x \leq
3200 M$, $-3200 M \leq y \leq 3200M$, and $0\leq z \leq 3200 M$ (we
used $\pi$-rotation symmetry and $z$-reflection symmetry). The
refinement levels were centered on the two BHs 
with half-widths of
1600, 800, 440, 220, 110, 55, 25, 10, 5, 2, and 0.75.
In the figures below we denote the
resolution of the coarsest grid by $h$. Our lowest-resolution runs had a coarsest resolution of $h=h_0 = 32M$. We increased
the resolution by successive factors of 1.2 for the higher-resolution
runs. Our standard choice, which we used for all long-term runs shown
below used a {\it medium} resolution of $h=h_0/1.2$. The highest-resolution run had $h=h_0/1.2^3$.
\begin{table}
\caption{Initial data parameters. $m_1$ and $m_2$ are the masses of
the two BHs, $D$ is the orbital separation, $\delta \Omega_{\rm orb}$ and
$\delta \dot r$ are the modifications to the 3.5PN orbital frequency
and inspiral required to reduce the eccentricity, $\Xi$ is a scale
factor (see text), and $r_H$ is the measured horizon radius.}
\label{table:ID}
\begin{ruledtabular}
\begin{tabular}{ll|ll}
$m_1/M$ & $0.5$ & $m_2/M$ & $0.5$ \\
$D/M$ & $20.00$ & $M\ \delta \Omega_{\rm orb}$ & $7.88515\times10^{-6}$ \\
$\delta \dot r$ & $-1.54103\times10^{-4}$ & $r_{\rm min}/M$ & 0.05 \\
 $r_{\rm max}/M $ & 0.25 & $\Xi$ & $800$ \\
$r_H/M$ & $0.484$
\end{tabular}
\end{ruledtabular}
\end{table}

\begin{figure}
  \includegraphics[width=.99\columnwidth]{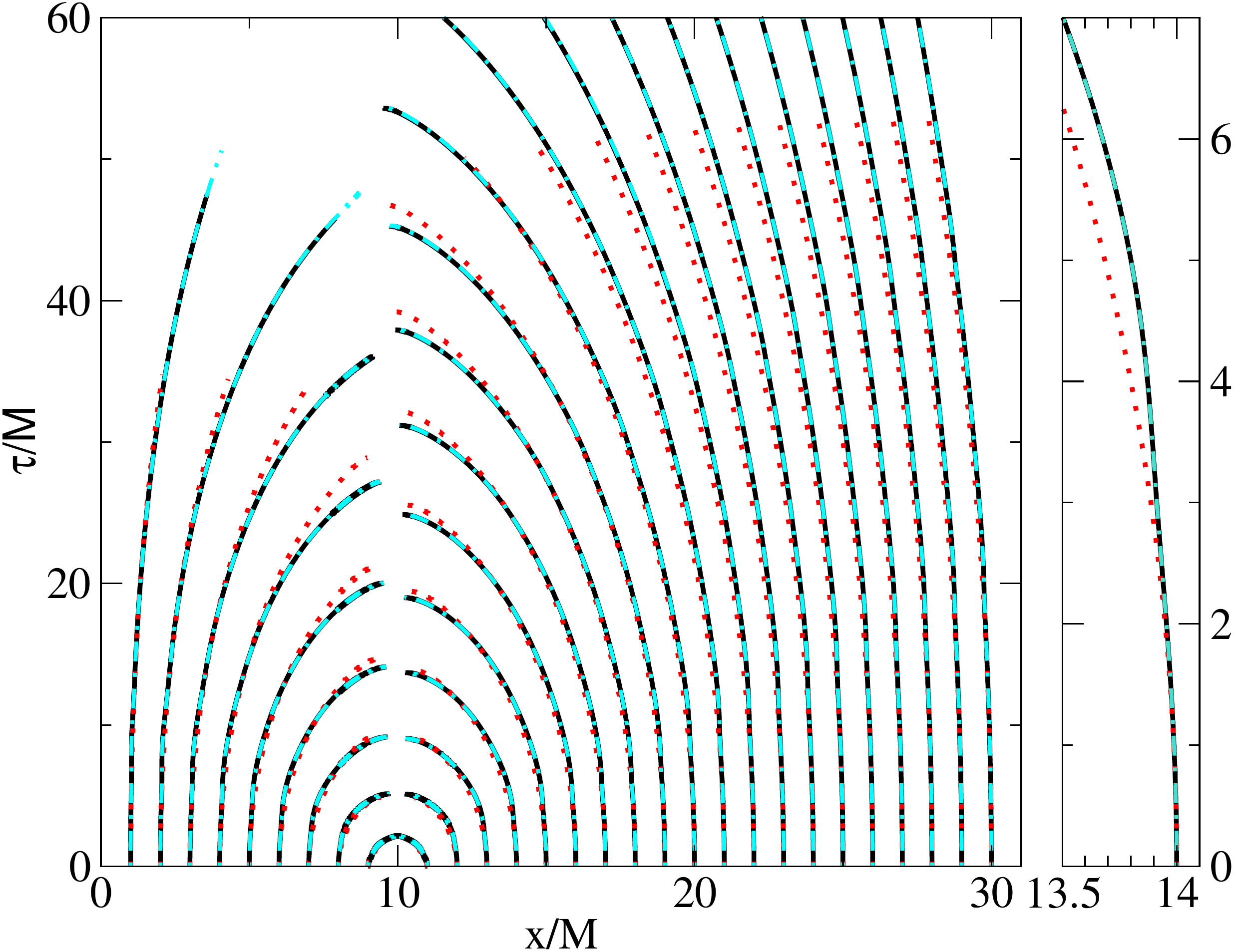}
  \caption{A set of timelike geodesics initially equally spaced in $x$ 
    and normal to the $t=0$ hypersurface (note that one of the BHs is
    centered at $x=10M$ at $t=0$). Here the proper time of each
    geodesic is plotted as a function of the geodesic's spatial
    position.
    The solid (black) curves correspond
    to geodesics evolved on the numerical spacetime using BSSN, the
    dot-dashed 
    (cyan) curve correspond to the geodesics evolved on the
    numerical spacetime using CCZ4, finally the dotted (red)
    curves are for geodesics evolved with the second-order analytical
    metric. The plot to the right zooms in on a typical geodesic near 
    the start of
    the simulation. Note how the numerical spacetime geodesics smoothly
    deviate from their analytical spacetime counterparts and there is
    no noticeable difference between the test particle trajectories
    for the BSSN and CCZ4 spacetimes at these early times.
     The rapid
    change in gauge near the start of the simulation is apparent in
    the smooth change in the geodesic seen at a proper time of about
  $\tau=3M$ in the inset.}\label{fig:geod}
\end{figure}

To demonstrate the smoothness of the transition from analytical
to numerical evolution we,
evolve a set of test particles using both the second-order analytic
metric and the numerical evolution of the second-order analytic data.
 Note that at
$t=0$, the 4-metrics associated with the two evolution schemes are
geometrically identical (i.e., they only differ by a coordinate
transformation implicit in using different choices for the lapse and
shift at $t=0$). However, because the evolution schemes are
different, the two spacetimes will have different effective stress
energies (i.e., $G_{ab}$ will differ for the two spacetimes) even at
$t=0$. Thus, even if the two spacetimes were initially in the same
coordinate system, higher-order time derivatives (third and higher) of the
geodesics will not agree. Thus, we only expect continuity of the {\it force}
acting on the geodesics as we transition from analytical to numerical
evolutions of the metric.
As shown in Fig.~\ref{fig:geod}, we do see a relatively smooth transition at
early times with the two sets of geodesics initially agreeing quite
well and then deviating more significantly at later times.
Importantly, this latter deviation is due to two effects, differences
in the later time coordinate systems and differences in the curvature.
Plots of the
same geodesics from the CCZ4 and BSSN evolutions are nearly identical.

Figure~\ref{fig:geod} provides evidence that the transition from
analytical to numerical evolutions is sufficiently smooth that no
sudden impulses  are imparted to timelike geodesics. However, we still need
to demonstrate that the subsequent dynamics are accurate. This will
require that the constraint violations do not significantly affect the
dynamics of the binary and that the binary remains quasicircular.

\subsection{Initial Hamiltonian and momentum constraint violations}

Evolutions of second-order analytical data
with BSSN were first performed in
Tichy~\cite{Reifenberger:2012yg}, where, like we do
here, they looked at
mass conservation, inspiral time, anomalous eccentricity, and
constraint violations (see also Ref.~\cite{Kelly:2009js}). The
conclusion there, as well as here, is that residual constraint
violations lead to relatively large errors in the subsequent dynamics.

Our expectation is that inaccuracies in the second-order analytic
metric will decrease as the binary separation is increased. To test
this assumption, we need a measure of the constraint violation that is
related to the dynamics of the binary.
Since we can interpret violations of the Hamiltonian constraint as an
unphysical matter field on our spacetime, a natural measure of the
degree of violation is the total amount of unphysical matter compared
to the total amount of physical mass (in this case, the total amount
of physical mass is $\approx 1M$). One subtlety we have to contend
with is that both positive and negative mass densities are dynamically
important, and there is no reason to expect their respective
contributions to the total error will cancel. Thus, we consider two 
measures of the unphysical mass given by
\begin{eqnarray}
  m_{\rm unphys.} &=& \frac{1}{16 \pi}\int {\cal H} \sqrt{\gamma} dv \,, \\
  m_{\rm abs.} &=& \frac{1}{16 \pi}\int |{\cal H}| \sqrt{\gamma} dv \,, 
\end{eqnarray}
where ${\cal H}$ is the Hamiltonian constraint violation and 
the integral is over the cube centered on the origin with side
length $400M$  excluding the interiors of the horizons. The former
gives the net unphysical mass in the spacetime, while the absolute
value in the latter ensures that positive and negative mass densities
do not cancel.

\begin{figure}
   \includegraphics[width=.99\columnwidth]{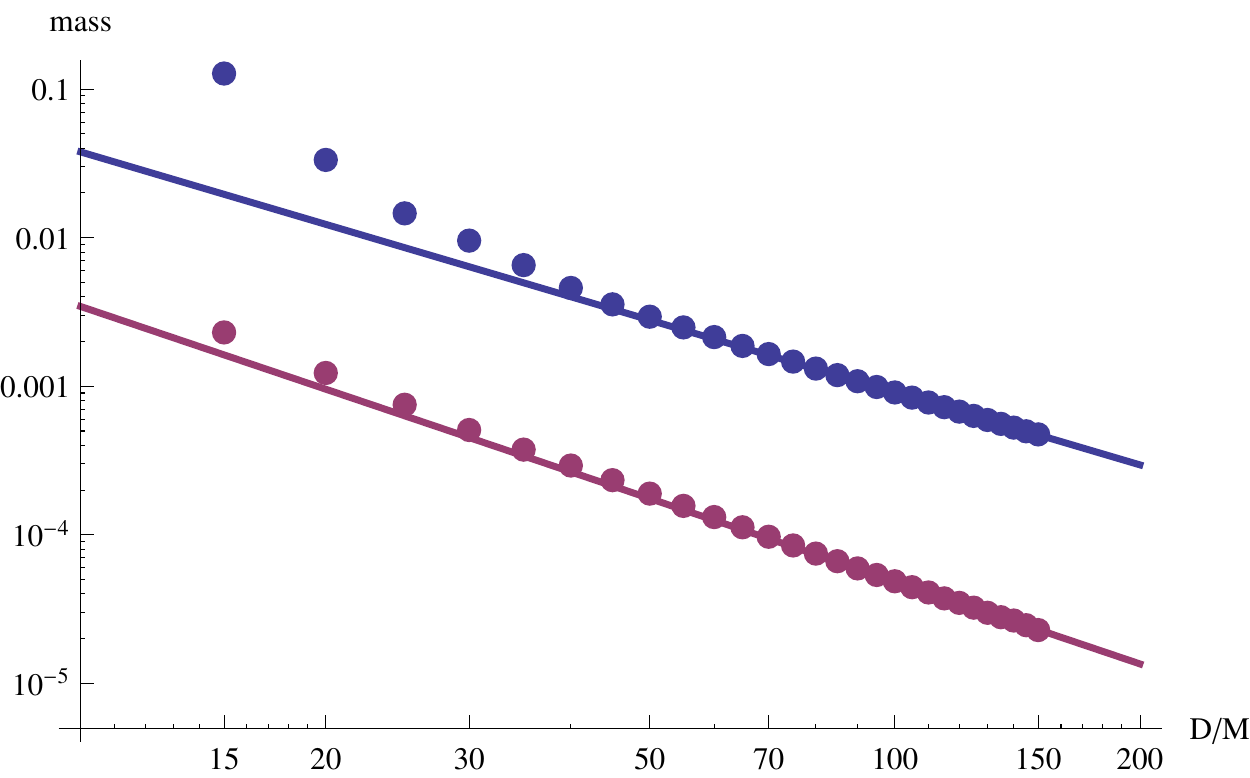}
   \caption{The unphysical mass of the binary versus binary separation
     $D$ for the second-order analytical data (i.e., initial data)
     as measured using the
     integrals $ m_{\rm unphys.}$ and $m_{\rm abs.}$.  The horizontal
     axis is in units of $D/M$, where $D$ is the separation of the
     binary. The larger mass
   is $m_{\rm abs.}$. Note that while a significant amount
 of unphysical matter is present, it is spread out such that only a
 fraction of it is absorbed by the BHs (see Fig.~\ref{fig:HC_V_PN}). Note also that there
 is a near equal amount of positive and negative mass (which is why
 $m_{\rm unphys.}$ is about a factor of 20 smaller than $ m_{\rm
 abs.}$). Rather than plotting $m_{\rm unphys.}$, we plot $-m_{\rm
 unphys.}$, since the net unphysical mass is actually negative.}\label{fig:unphysical}
\end{figure}

We show both of these measures of unphysical mass versus separation in
Fig.~\ref{fig:unphysical}.
Note that there
is a near equal amount of positive and negative mass (which is why
$m_{\rm unphys.}$ is about a factor of 20 smaller than $ m_{\rm
abs.}$). The magnitude of $m_{\rm unphys.}$ decreases with binary separation
as roughly $(D/M)^{-1.8}$, while $ m_{\rm abs.}$ decreases at the rate of
$(D/M)^{-1.6}$. The masses were calculated at $t=0$ for three
resolutions. In all cases, the truncation errors for the highest
resolution corresponded to an uncertainty in the second or higher significant
digit in the mass.

At a separation of $D=20M$, we find that $m_{\rm unphys.}=0.001M$,
while $m_{\rm abs.}=0.033M$
(Bowen-York data for a $D=20M$ binary solved using the {\sc
TwoPunctures}~\cite{Ansorg:2004ds} code with
$40^3$ collocation points gives $|m_{\rm abs.}| < 7\times10^{-7}M$).
Note that $m_{\rm abs.}$ increases much more rapidly than
the
power-law prediction with decreasing
separation for $D<25M$, but $m_{\rm unphys.}$ is only $30\%$ larger
than the power-law prediction at $D=20M$.
It is important to note
that the BHs will not absorb this much mass, as their cross sections
are
quite small.

An important result from Fig.~\ref{fig:unphysical} is that while the
unphysical mass tends to zero at infinite binary separation, for
practical purposes, it is never {\it small} in the regime where we
would use NR evolutions. Thus, we need a way of
removing the unphysical mass from the system.

We also examine how the quantity of unphysical matter ($m_{\rm abs.}$)
depends on the locations of the inner/near buffer zones. For our runs,
we used the transition parameters of
Ref.~\cite{JohnsonMcDaniel:2009dq}, which were optimized for a
separation of $D=10M$. Using these parameters, we find that
$m_{\rm abs.}$ is $0.033M$. By optimizing the parameters to reduce
$m_{\rm abs.}$, we reduce this by only $4\%$ to $0.032M$.
Interestingly, while $m_{\rm abs.}$ is reduced, the constraint
violations are more concentrated near the two BHs, thus allowing
for more absorption of constraint-violating matter by the BHs.
Importantly, the constraint violations cannot be significantly reduced
by moving the locations of the zone boundaries, since they are fixed
analytic functions of the masses and separation delimiting overlapping
regions of validity for different metric approximations.

\begin{figure*}[!t]
  \includegraphics[width=0.475\textwidth]{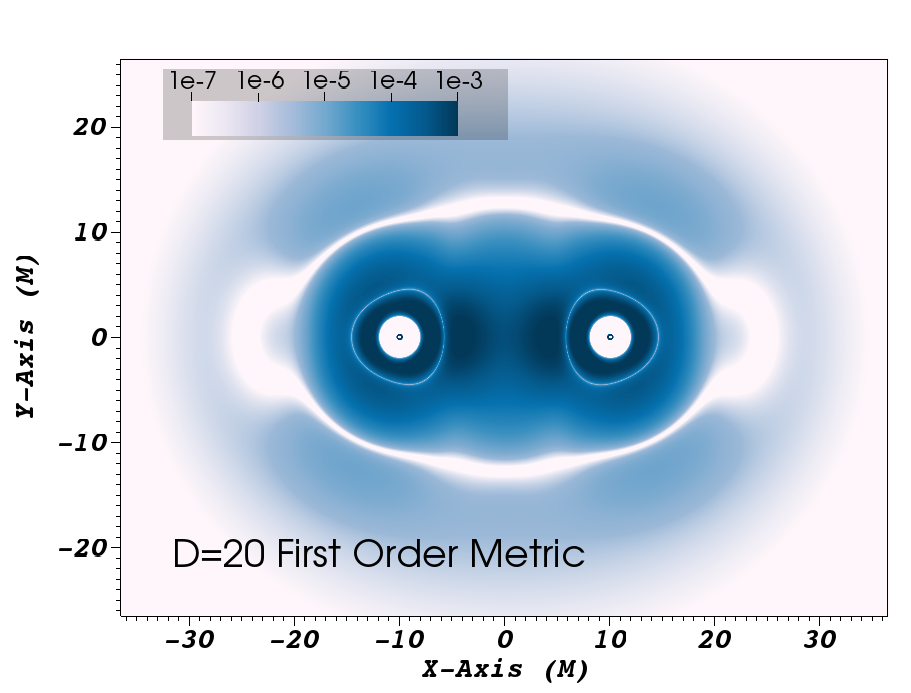}
  \includegraphics[width=0.48\textwidth]{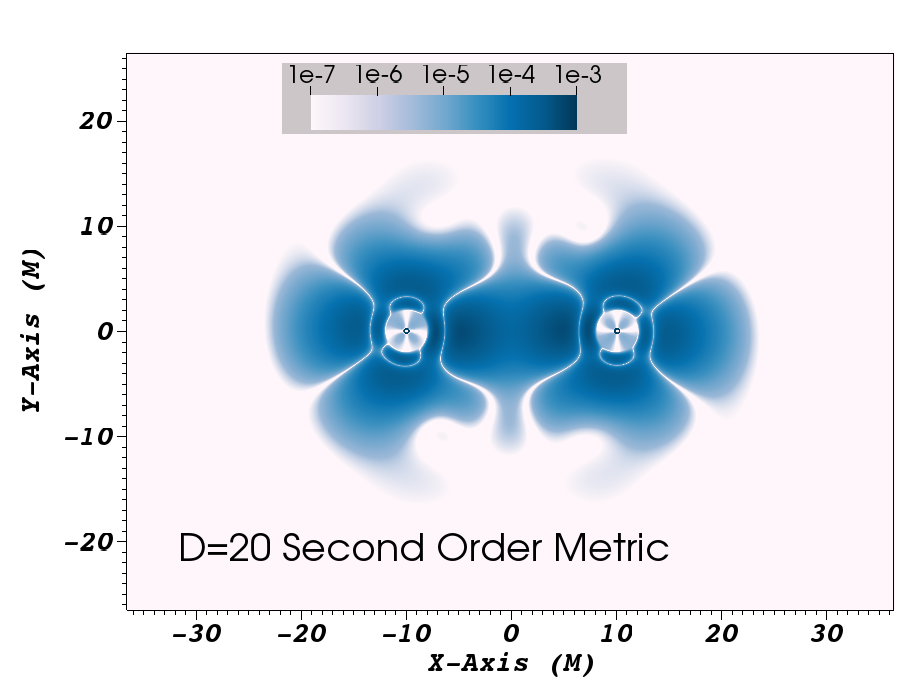}
  
  \includegraphics[width=0.49\textwidth]{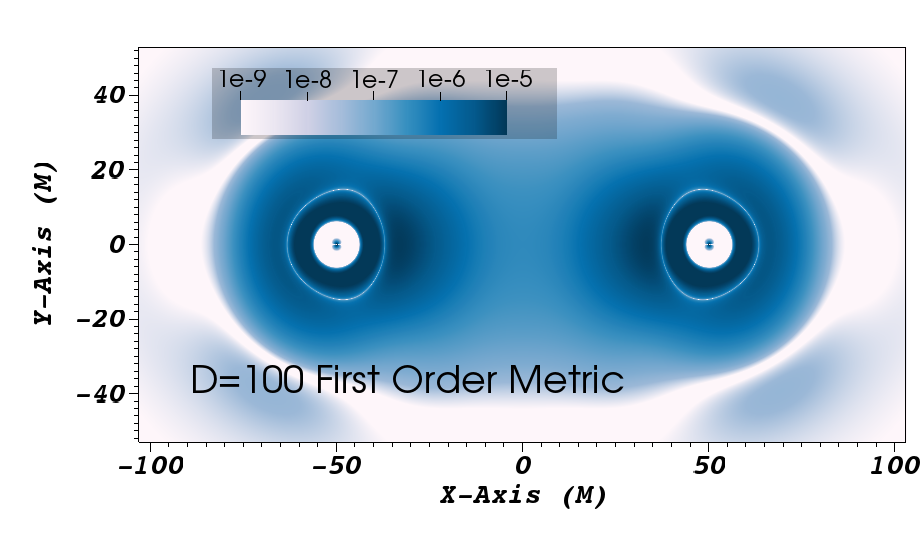}
  \includegraphics[width=0.495\textwidth]{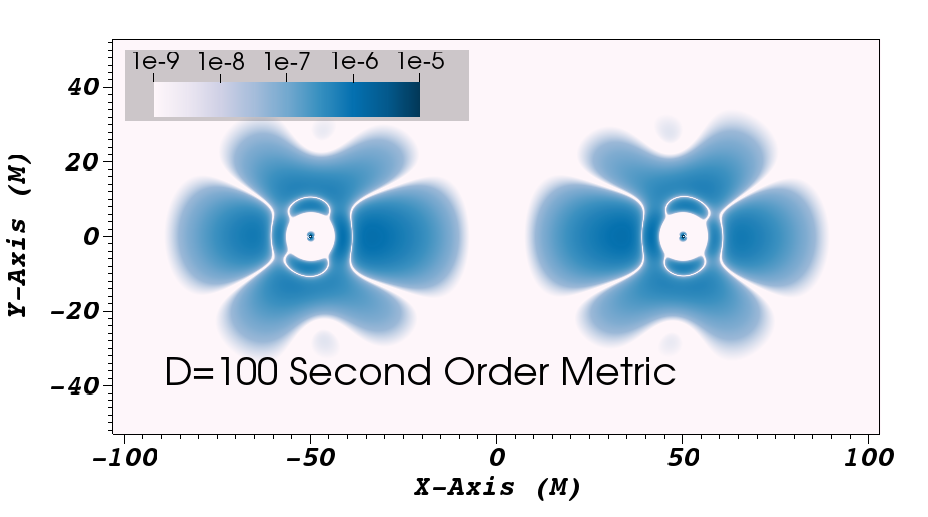}
  \caption{The Hamiltonian constraint violation in the vicinity of
  the two BHs in the binary for binary separations of $D=20M$ (top)
and $D=100M$ (bottom). First-order metrics are shown to the left,
and second-order metrics to the right. The very small white circles
at the centers of the BHs are the horizons.
}\label{fig:HC_V_PN}
\end{figure*}

It is important to determine not only how much unphysical matter is
present, but also where it is located.
To this end, we plot the Hamiltonian constraint violations on the
equatorial plane for BHBs at separations of $D=20M$ and $D=100M$ using both the
standard second-order analytical metric described in Sec.~\ref{sec:ID} and 
the first-order version.
The main difference between the two is described in
Table~\ref{tab:orders}.
The Hamiltonian constraint violations show a clear improvement as we
switch from the first-order to the second-order analytical metric,
indicating that it is the low PN order which dominates the error.
Perhaps unexpectedly, even at a separation of $D=100M$, the first-order metric has $m_{\rm
abs.} = 0.02M$, while the second-order metric has $m_{\rm abs.} =
0.001M$. Examining Fig.~\ref{fig:HC_V_PN}, we see that the constraint violations
are concentrated in extended {\it clouds} well outside the horizons 
in the buffer zone between
the inner and near zones. 
Most importantly, the first-order metric shows a strong shell
of high constraint violation surrounding the two BHs. The second-order
metric, on the other hand, has a lower-amplitude, more diffuse cloud of
constraint violation that is less likely to be absorbed.

\subsection{Effects of the Hamiltonian and momentum constraint violations}

In this section, we examine how the numerical evolution scheme can
compound  or mitigate issues associated with initial constraint
violations. To this end, we evolved the second-order analytical data
using both the BSSN formulation~\cite{Nakamura87, Shibata95,
Baumgarte99} and the constraint-damping CCZ4 approach.

Our initial explorations of the dynamics of the second-order
analytical data were based on the BSSN and CCZ4 systems. As shown in
this section, we find that the CCZ4 is uniformly better than BSSN in
evolving data with nontrivial constraint violations.

\begin{figure}
  \includegraphics[width=.99\columnwidth]{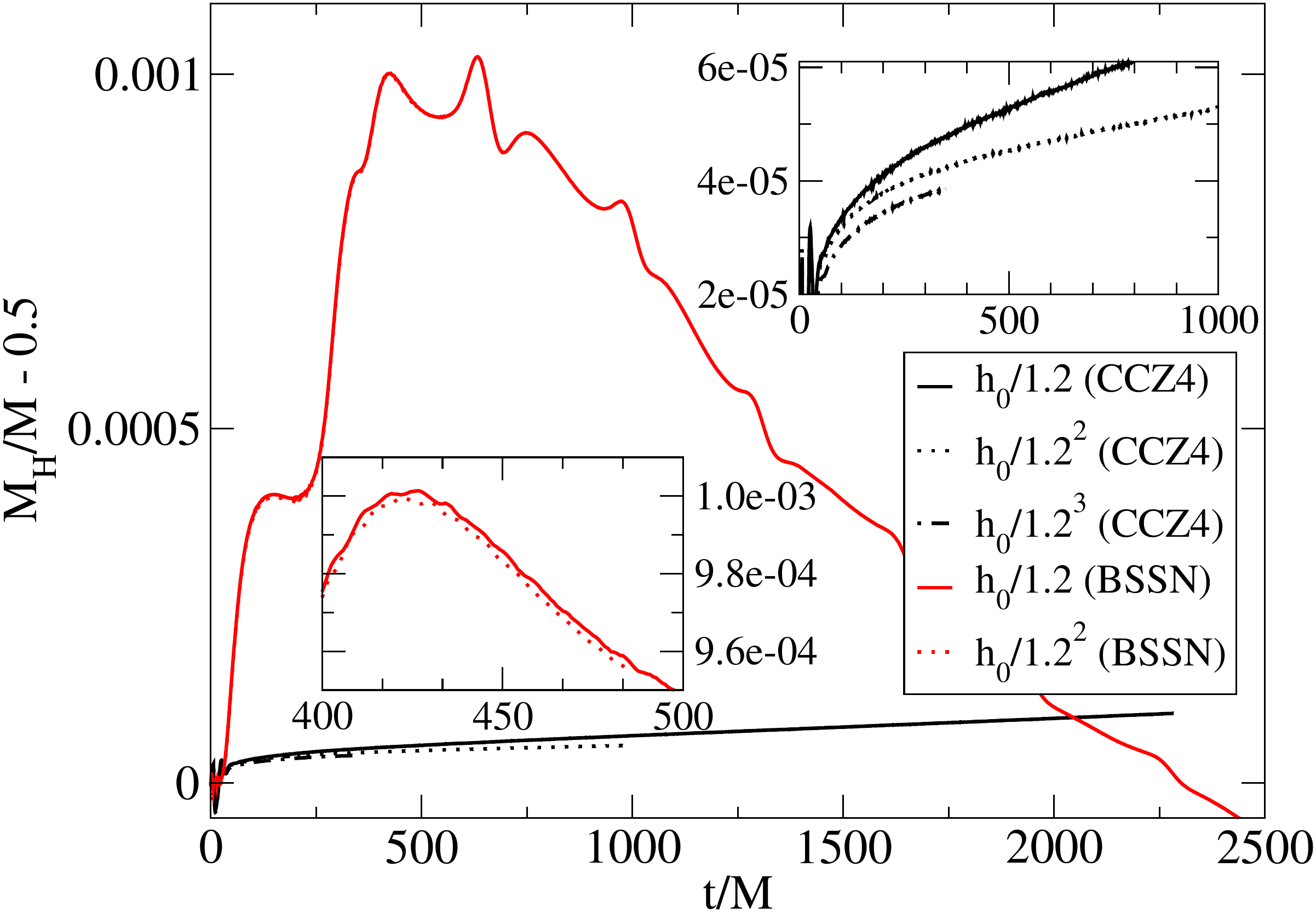}
  \caption{The individual apparent horizon masses for a BHB with initial
    separation of 20M for both CCZ4 (black) and BSSN (red) evolutions of the
second-order analytical data.  The BSSN curves show very large
oscillations,
while the CCZ4  curves show a much smaller linear growth.
The large mass
oscillations in the BSSN run are due to absorption of constraint
violations. Note how the effects changing the evolution system from
BSSN to CCZ4 are much larger than truncation error effects.}\label{fig:BSSN_mass_conservation}
\end{figure}

One of the most important differences between the BSSN and CCZ4 evolutions is in
the horizon mass conservation.
As shown in
Fig.~\ref{fig:BSSN_mass_conservation}, the mass conservation of the
BHs was relatively poor for BSSN and substantially better for CCZ4. In
the figure, we see that the BSSN run showed an initial increase in mass
of $10^{-3}M$, followed by a mass loss of similar magnitude.  While a
change of 2 parts in 1000 may seem small, the effect of this mass
change on the orbital trajectory is quite large.

To determine the cause of the lack of conservation of the (apparent) horizon mass, we
compare the time derivative of the horizon mass
$dM_H/dt = dM_{1,2}/dt$ ($M_1=M_2$ by symmetry) with the average value
of the constraint on the horizons ${\cal H}_H$  and the flux of
constraint violation into the horizon ${\cal C}_H$ (since the spacetime
around the two horizons is identical by symmetry, we only plot
the constraint violation averaged over one of the BHs). 
We define ${\cal H}_H$ and ${\cal C}_H$
as
\begin{eqnarray}
  {\cal H}_H =  \frac{\oint {\cal H} \sqrt{\sigma} dAdB}{\oint 
  \sqrt{\sigma} dAdB} \,,\\
  {\cal C}_H =  - \oint {\cal C}^i n_i \sqrt{\sigma} dAdB \,,
\end{eqnarray}
where ${\cal H}$ is the Hamiltonian constraint violation, ${\cal C}^i$
is the momentum constraint violation, $n_i$ is the unit (outward) normal to the horizon,
$\sqrt{\sigma} dAdB$ is the proper area element on the horizon,
and the integrals are performed over the surface of the horizon.

\begin{figure}
  \includegraphics[width=\columnwidth]{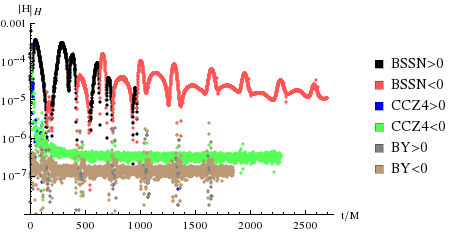}
  \includegraphics[width=\columnwidth]{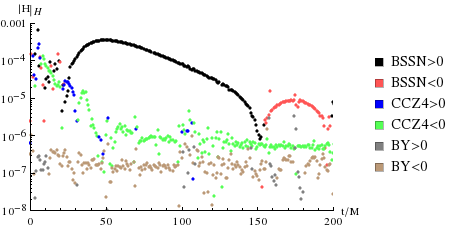}
  \caption{The Hamiltonian constraint averaged over the horizon as a
  function of time. This linear log plot shows the absolute value of
the constraint violation for the BSSN and CCZ4 evolutions of the
second-order analytic metric, as well as the evolution of
Bowen-York data (BY) using CCZ4. The BSSN,
CCZ4, and BY  data points are colored according to the sign of the constraint
violation.  The lower plot shows the
early-time behavior.
}\label{fig:H_H}
\end{figure}

In Fig.~\ref{fig:H_H}, we show the constraint
violation averaged over the individual horizons for both BSSN and CCZ4. A large positive
violation is observed at early times for BSSN, which is followed by a
negative constraint violation. This links well with the initial
increase in horizon mass for BSSN, which is followed by a later-time
decrease.
The CCZ4 constraints are a factor of 100 smaller and do not
appear to be correlated with the CCZ4 horizon mass.
Because of these results, all of our long-term
simulations used CCZ4.

\begin{figure}
    \includegraphics[width=.98\columnwidth]{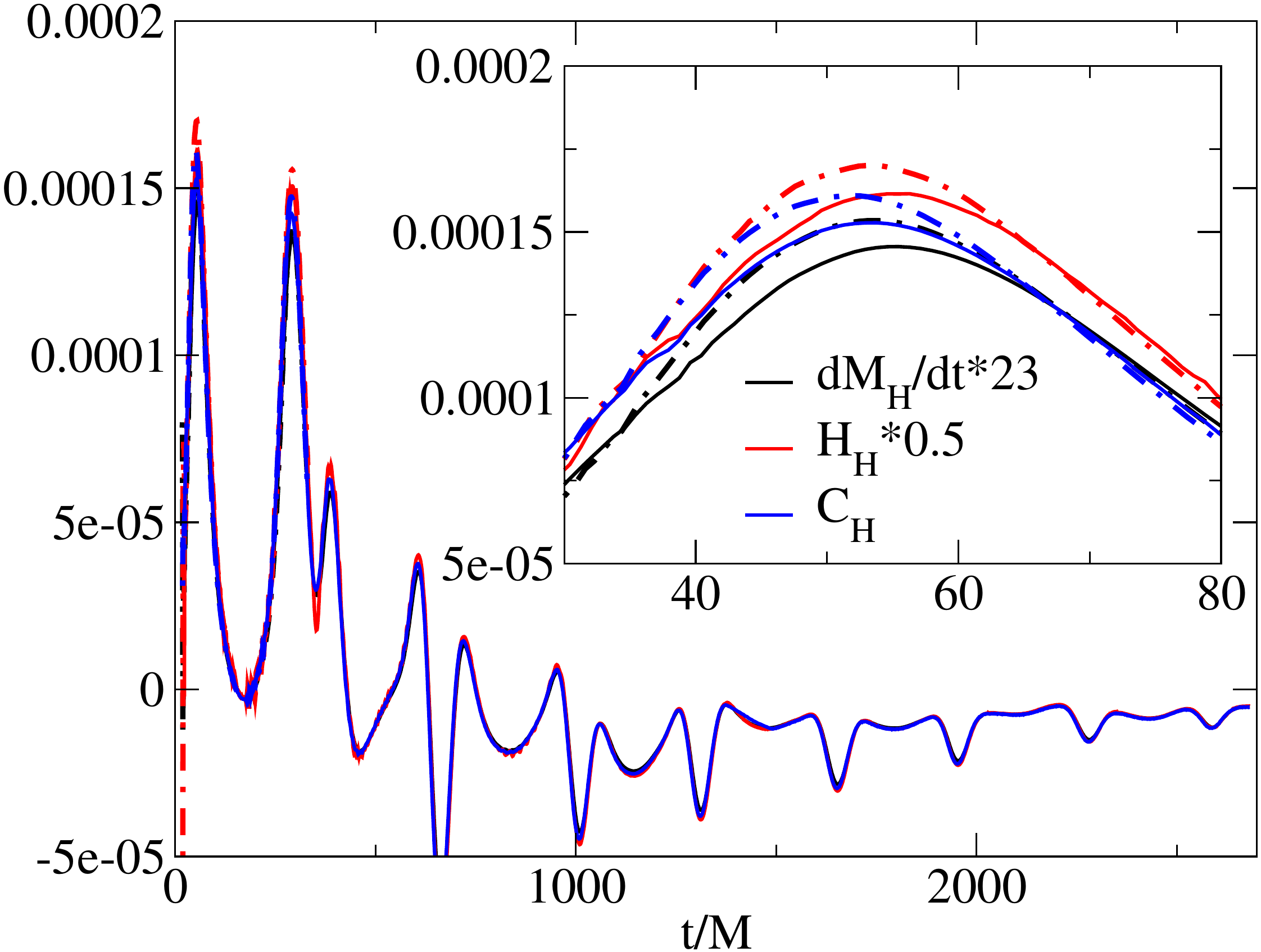}
    \caption{The Hamiltonian constraint averaged over the horizon, the flux of the momentum constraint
        violations through the horizon, and
        the time derivative of the horizon mass for the BSSN
        simulation.
        The Hamiltonian and change of rate of the horizon mass have
        been multiplied by constant positive factors.
Note the near-perfect correlation of horizon mass change
        and constraint violation on the horizon. In the figure, solid
        curves correspond to a resolution of $h_0/1.2$, while dotted
        curves correspond to a resolution of $h_0/1.2^2$.
      } \label{fig:H_H_cmp_mdot}
\end{figure}

\begin{figure}
    \includegraphics[width=.99\columnwidth]{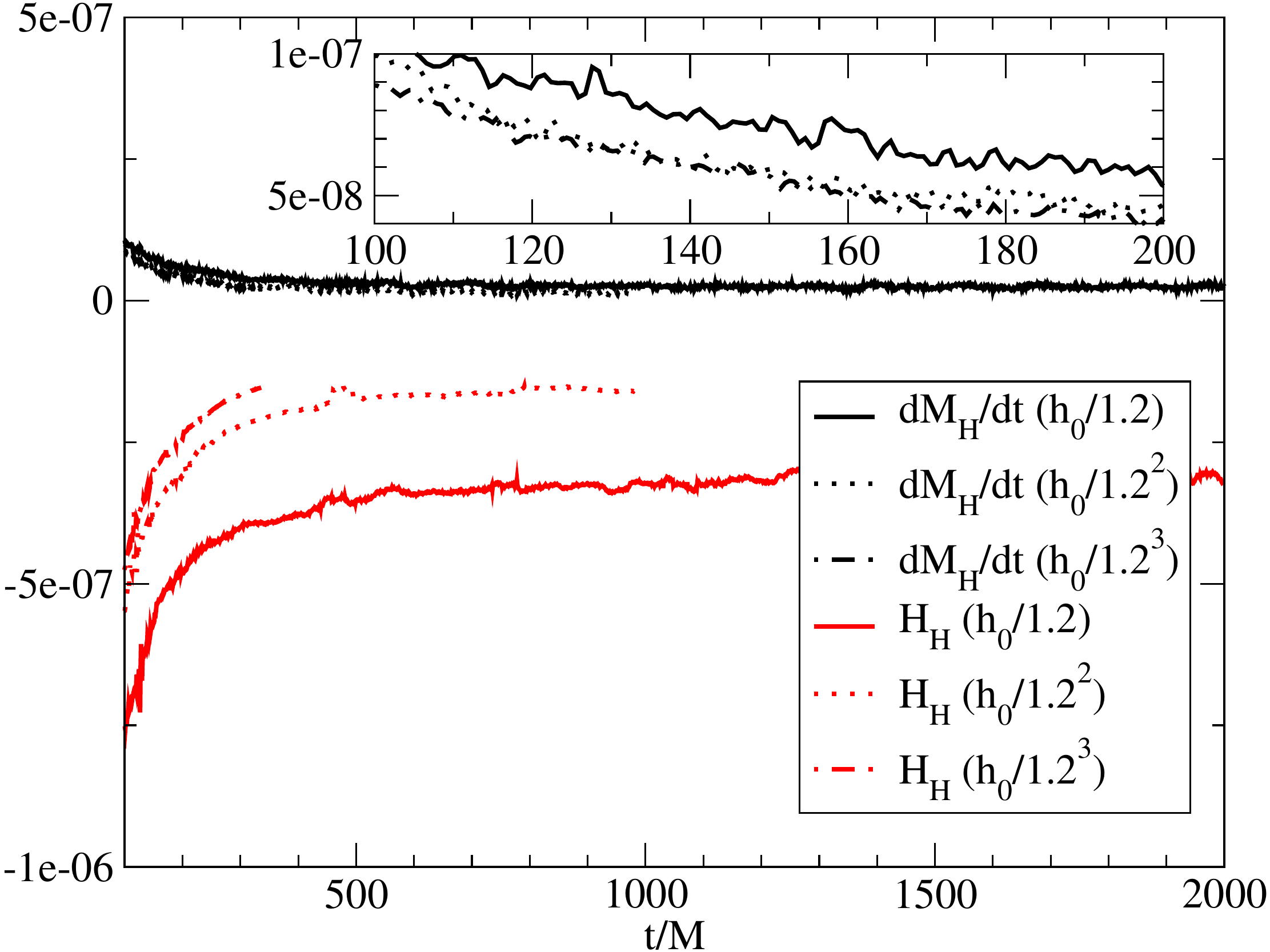}
    \caption{The Hamiltonian constraint averaged over the horizon and
        the time derivative of the horizon mass for the CCZ4
        simulation. Note how unlike in Fig.~\ref{fig:H_H_cmp_mdot},
        these two are not correlated. Here both the constraint
        violation and $\dot m$ appear to be converging to a very small
        value but from opposite directions.
      } \label{fig:H_H_cmp_mdot_ccz4}
\end{figure}

Aside from an
overall positive (constant) numerical factor, plots of 
${\cal H}_H$, ${\cal C}_H$, and $dM_{H}/dt$ 
are nearly identical for BSSN (see  Fig.~\ref{fig:H_H_cmp_mdot}).
This means that all three are strongly correlating (i.e., $dM_{H}/dt
\propto {\cal C}_H \propto {\cal H}_H$).
This provides a compelling
argument that it is the constraint violations that cause the
horizon masses to fluctuate. On the other hand, for CCZ4, there is no
compelling correlation between $dM_{H}/dt$ and the (much smaller) constraint
violations (see Fig.~\ref{fig:H_H_cmp_mdot_ccz4}).

\begin{figure}
  \includegraphics[width=.98\columnwidth]{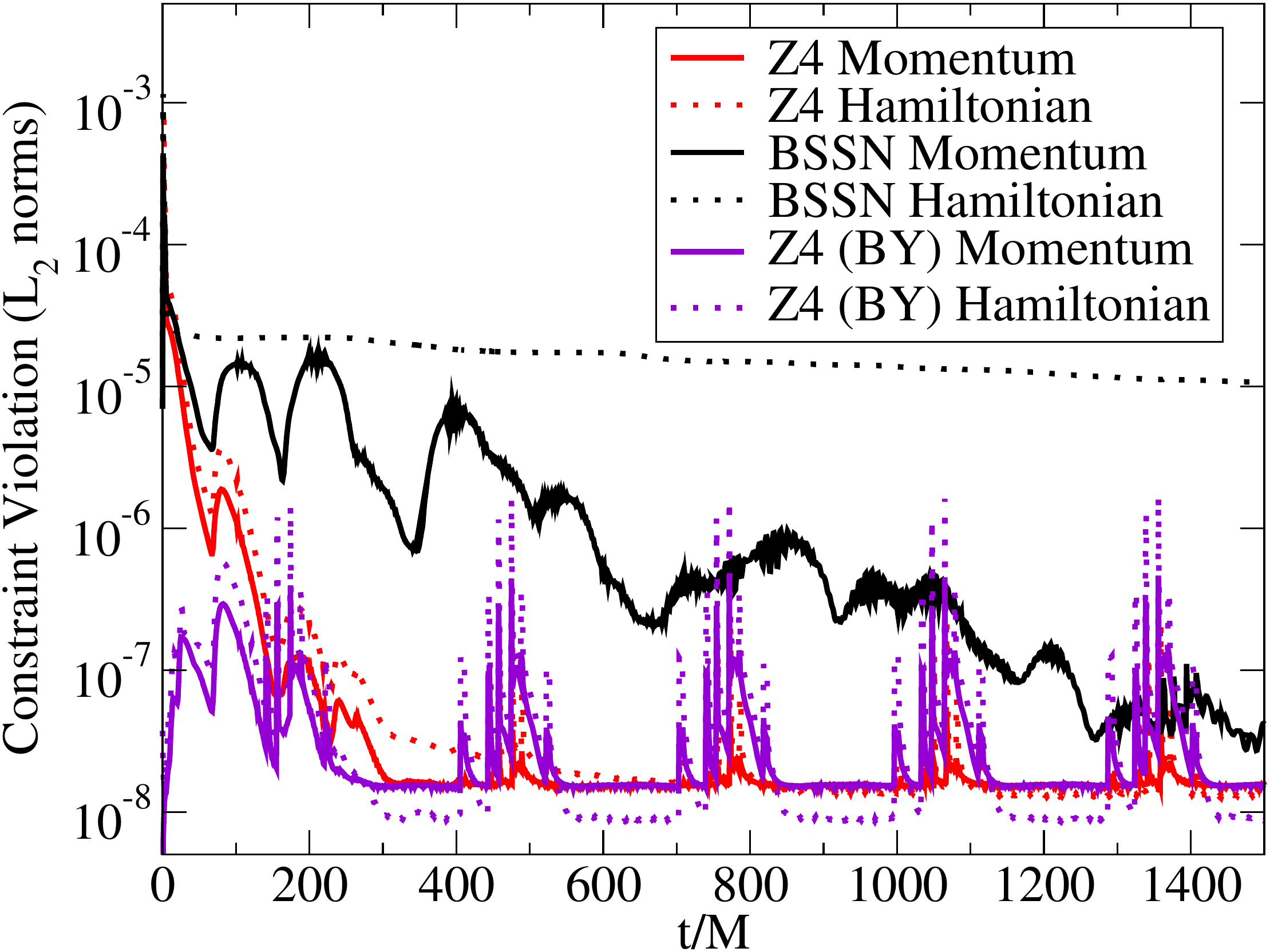}
  \caption{Time evolution of the $L^2$ norms of the constraint violations
for BSSN and CCZ4 evolutions. Note how the BSSN Hamiltonian
constraint remains relatively high while the CCZ4 constraints quickly
fall to about $10^{-7}$. Also shown are the $L^2$ norms of the constraint
violations for an evolution of Bowen-York data with CCZ4. The
Bowen-York data leads to constraint violations that are on average
a factor of 2 below constraint violations for the new data.
}\label{fig:HC_cmp1}
\end{figure}

Finally, we examine how the constraint violations in the bulk of
the simulation domain behave with time.
As shown in Fig.~\ref{fig:HC_cmp1}, the $L^2$ norms of the constraint
violations for CCZ4 and BSSN evolutions behave quite differently (here
we restrict the $L^2$ norm to the volume inside a ball of radius $30M$
and outside the two horizons so that the norm is dominated by
constraint violations relatively close to the binary).  The
CCZ4 constraints fall to a much lower level (about a factor of 1000
smaller for the Hamiltonian and a factor of 100 for the momentum) than
BSSN. For comparison purposes, we also performed equivalent evolutions with
standard Bowen-York initial data~\cite{Bowen:1980yu}.

\begin{figure}
  \includegraphics[width=.99\columnwidth]{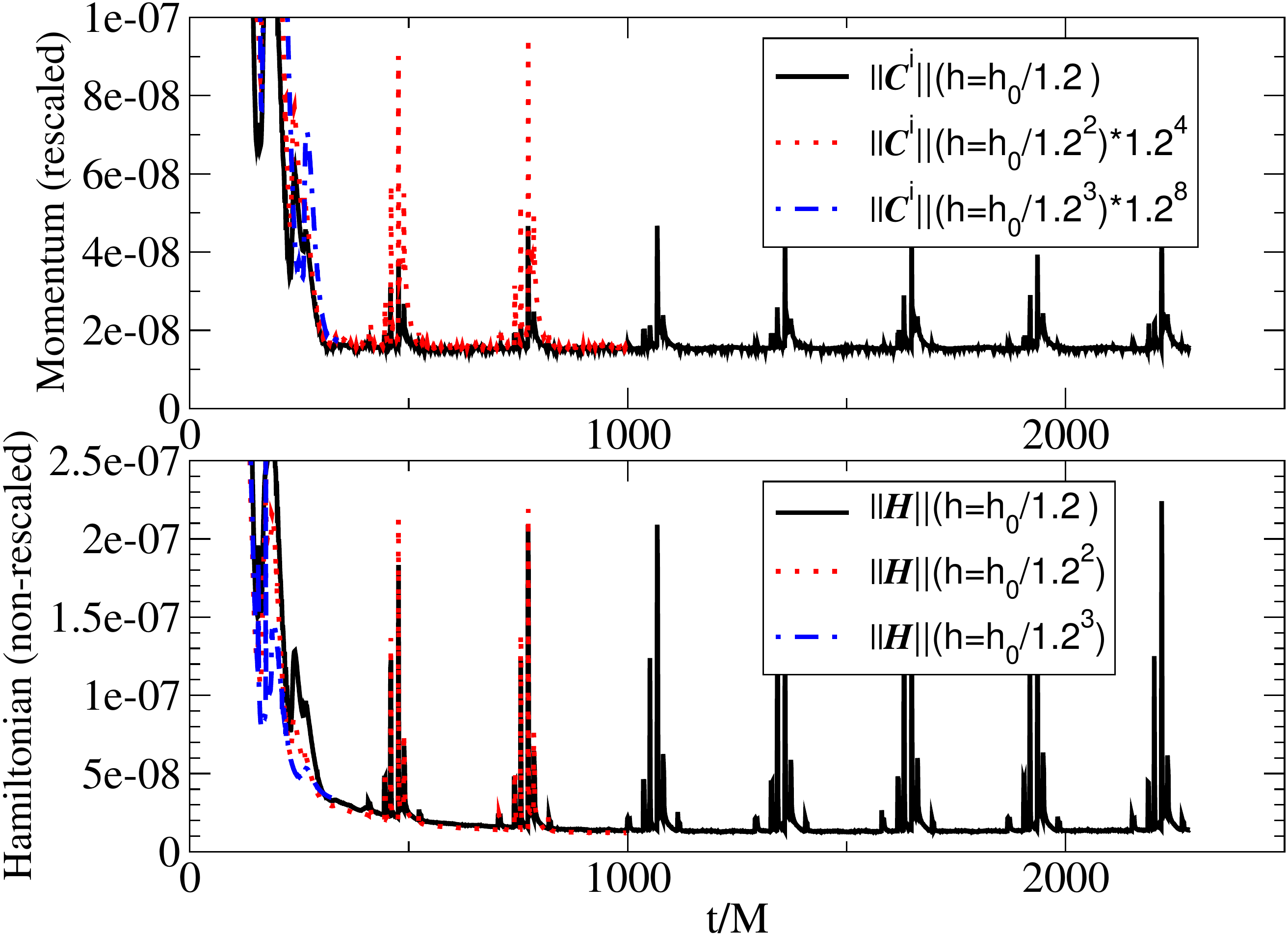}
  \caption{The $L^2$ norm of Hamiltonian constraint violations and
    the $L^2$ norm of the  Euclidean 
    magnitude of the momentum constraint violations.
The momentum constraints  have been rescaled by a factor of
$(h_0/h)^4$, where $h$ is the base resolution of a particular run
and $h_0$ is the
base resolution of the coarsest simulation. Note that the
Hamiltonian violations have not been rescaled. The spikes occurring
roughly every 350M are due to the high-frequency initial gauge wave
reflections off the mesh refinement
boundaries. Here too, the $L^2$ norms have been restricted to the
interior of a sphere of radius $30M$ about the origin and outside
the two horizons
}\label{fig:Z4_const}
\end{figure}

As shown in Fig.~\ref{fig:Z4_const}, we see clear
convergence of the $L^2$ norm of the momentum constraint violation to
zero for CCZ4, while
the Hamiltonian constraint violation, though small, seems to bottom out
at about $10^{-8}$. The Hamiltonian  constraint for an equivalent Bowen-York
run bottoms out at roughly half this value. For this convergence
check, we
only ran the highest-resolution run for 350M due to computational
costs.

\begin{figure*}
  \includegraphics[width=.49\textwidth]{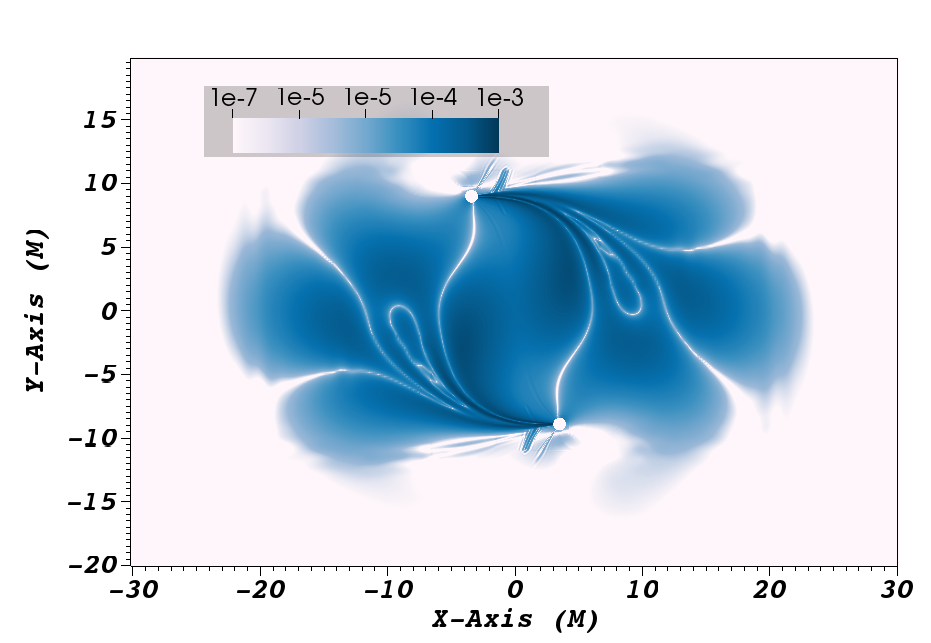}
  \includegraphics[width=.487\textwidth]{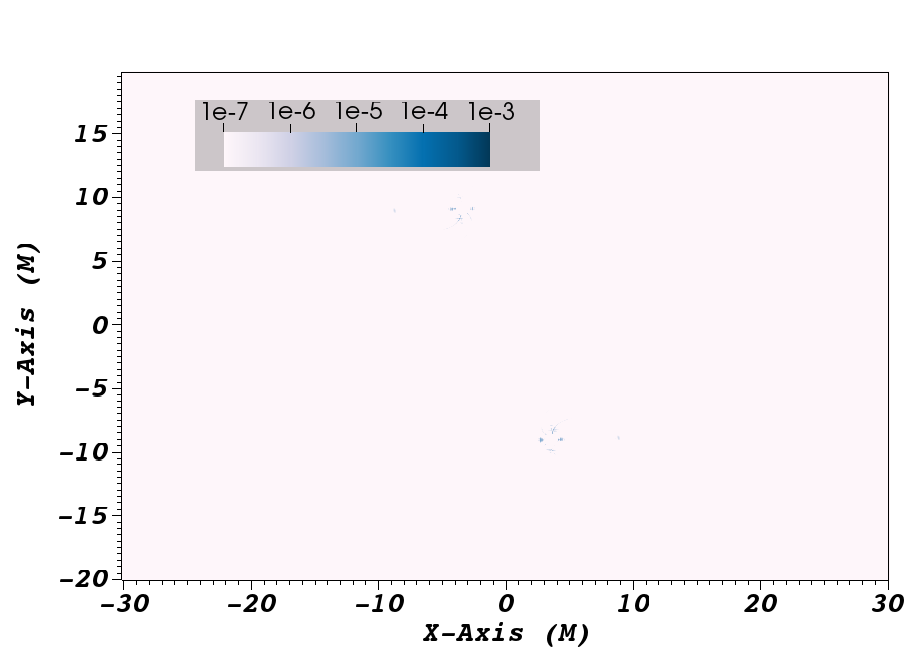}
  \caption{The Hamiltonian constraint on the $xy$ plane near the
two BHs at $t=200M$ for the BSSN (left) and CCZ4
(right) evolutions of the second-order analytical data starting at
a separation of $D=20M$.  The scale is logarithmic and goes from $10^{-7}$ to
$10^{-3}$. The absolute value of the constraint violations for the BSSN
simulation match closely the violations on the initial slice, while the CCZ4
violations are three orders of magnitude smaller.  In each plot the interiors of the
BHs have been masked out. The constraint violations at $t=0$ are
given in the top right plot in Fig.~\ref{fig:HC_V_PN}.
}\label{fig:HC_cmp2}
\end{figure*}

The amount of unphysical mass that the BHs can absorb
depends not only on the amount of unphysical matter, but also on the 
{\it dynamics} of the unphysical matter. For BSSN, the constraint
violations largely stay in place (and can therefore be accreted) due
to the presence of a zero-speed constraint mode in BSSN~\cite{Bernuzzi:2009ex},
while
for CCZ4 the constraints quickly leave the vicinity of the BHs. The very different
behavior of BSSN and CCZ4 is shown in Fig.~\ref{fig:HC_cmp2}.

\begin{figure}
  \includegraphics[width=.99\columnwidth]{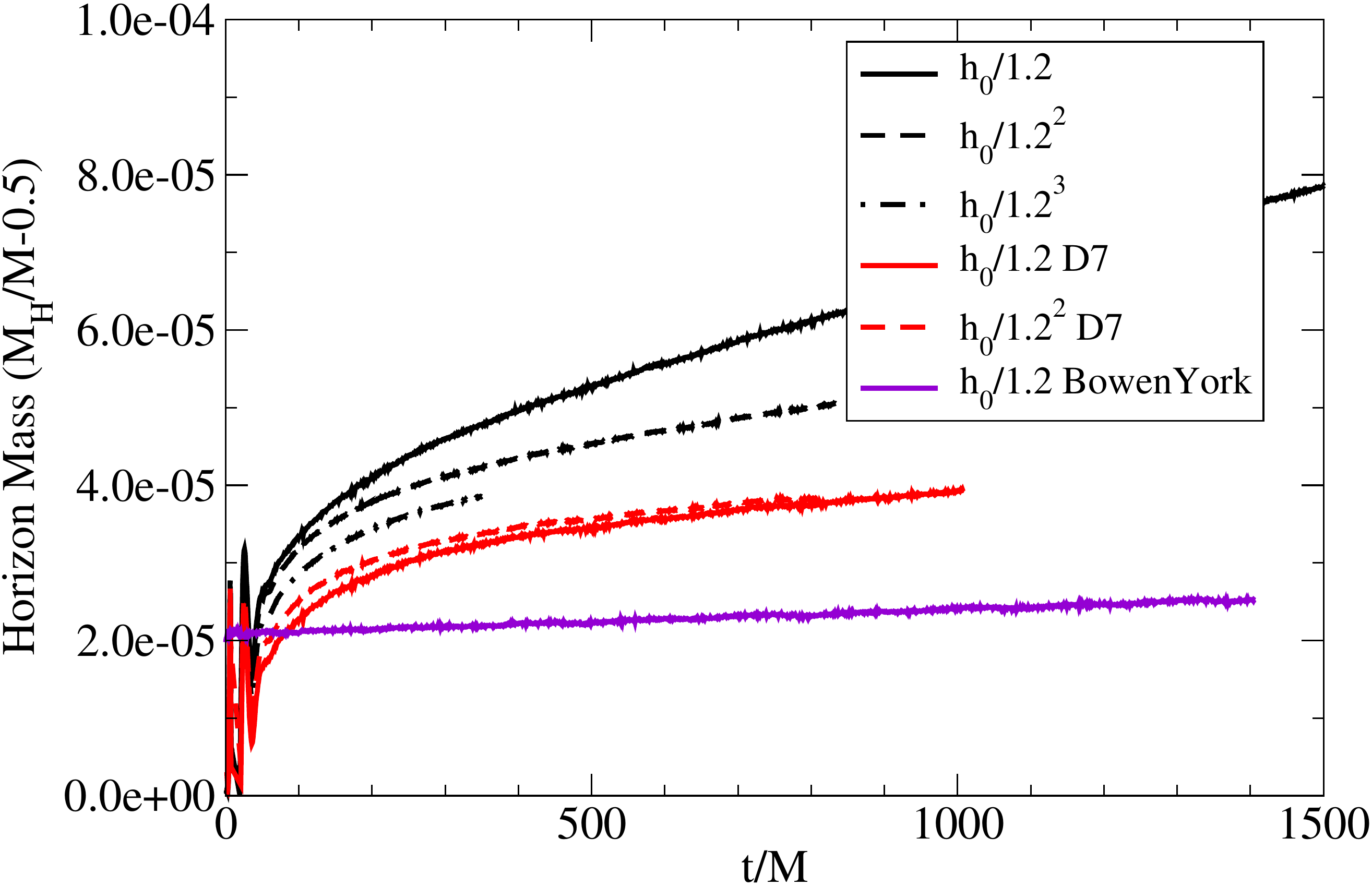}
  \caption{The mass of the individual horizons versus time for the
    CCZ4 simulations using the standard fifth-order dissipation,
    seventh-order dissipation, and fifth-order dissipation of
    Bowen-York data.
The linear trend in the mass, while 
converging to a small value, is substantially larger than 
that for Bowen-York.}\label{fig:Z4_mass}
\end{figure}

\begin{figure}
  \includegraphics[width=.98\columnwidth]{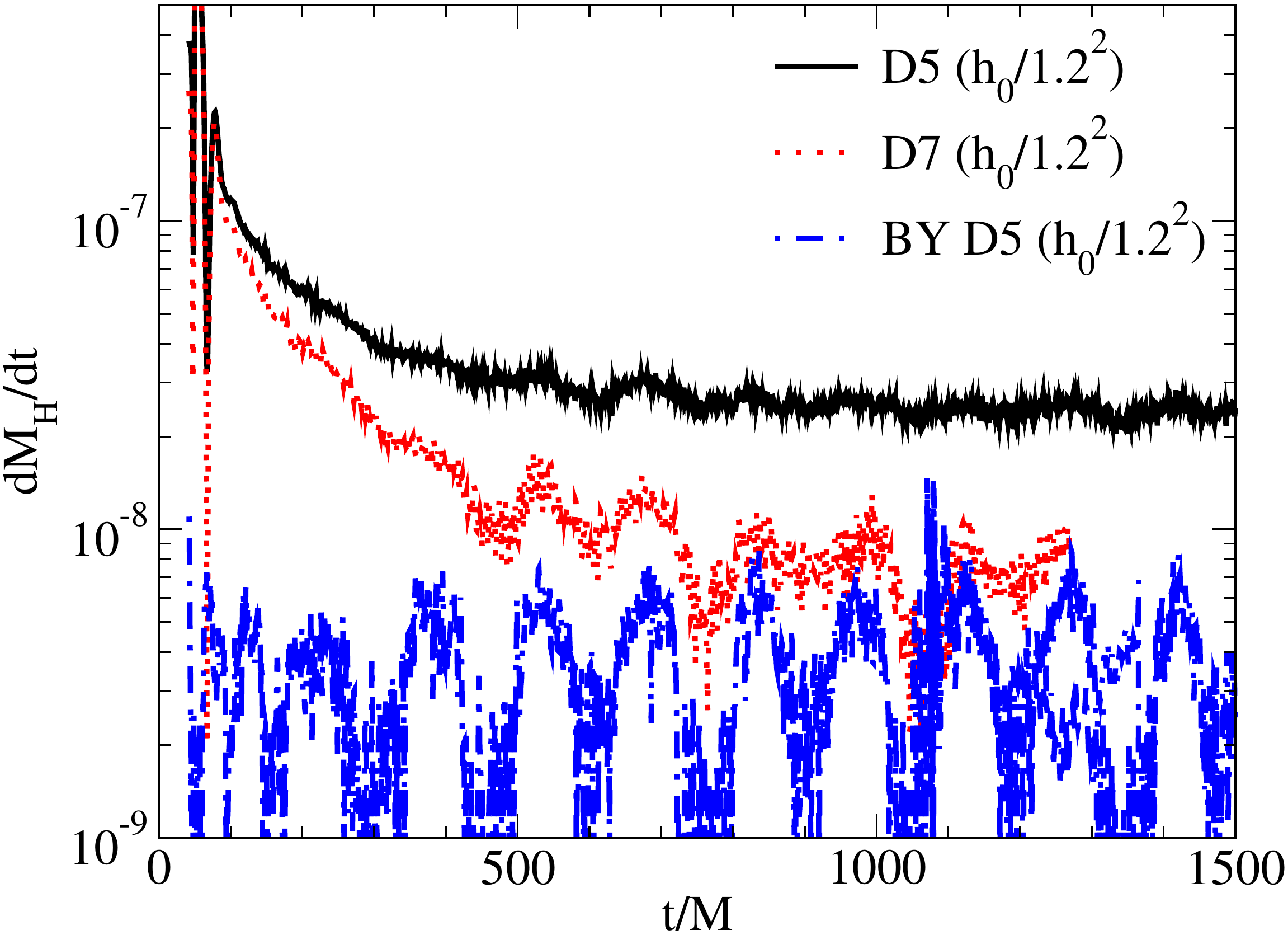}
  \caption{The rate of horizon mass increase versus time. Here D5
indicates that fifth-order dissipation was used, while D7 indicates
that seventh-order was used.}\label{fig:Z4_mass2}
\end{figure}

The overall efficacy of using CCZ4 to drive the constraint violations
to zero can be measured by examining in detail how well the horizon
masses are conserved.  As shown in Figs.~\ref{fig:Z4_mass} and
\ref{fig:Z4_mass2}, there is a relatively strong linear trend in the
mass that, while converging to a small value, is substantially larger than the Bowen-York result.
Here we also see a significant advantage to using
 higher-order dissipation.  Note that even with the
highest-resolution runs, the horizon mass increase is an order of
magnitude larger than
 for Bowen-York data evolved with CCZ4. 
Since the Bowen-York data were evolved with the same evolution
system and grid structure, it appears that there are peculiarities
associated with the analytic initial data driving the mass increase
(note, as seen in Fig.~\ref{fig:H_H_cmp_mdot_ccz4}, absorption of constraint violation
seems not to be the cause of this mass increase).
One possibility which we have not explored in detail here is that the
methods used to {\it stuff} the BHs may be affecting the mass
conservation. The other candidate would be residual constraint
violations. Even for our highest-resolution run, the average
constraint violation on the horizon surface at late times was 50\%
larger than for Bowen-York (the global constraint violations
were, on average, an order of magnitude larger, as shown in
Fig.~\ref{fig:HC_cmp1}).
As observed in Ref.~\cite{Zilhao:2014ida}, differences in accuracy of the spacetime at this
level will likely not be important for MHD simulations.

\subsection{Eccentricity}
\begin{figure}
  \includegraphics[width=.98\columnwidth]{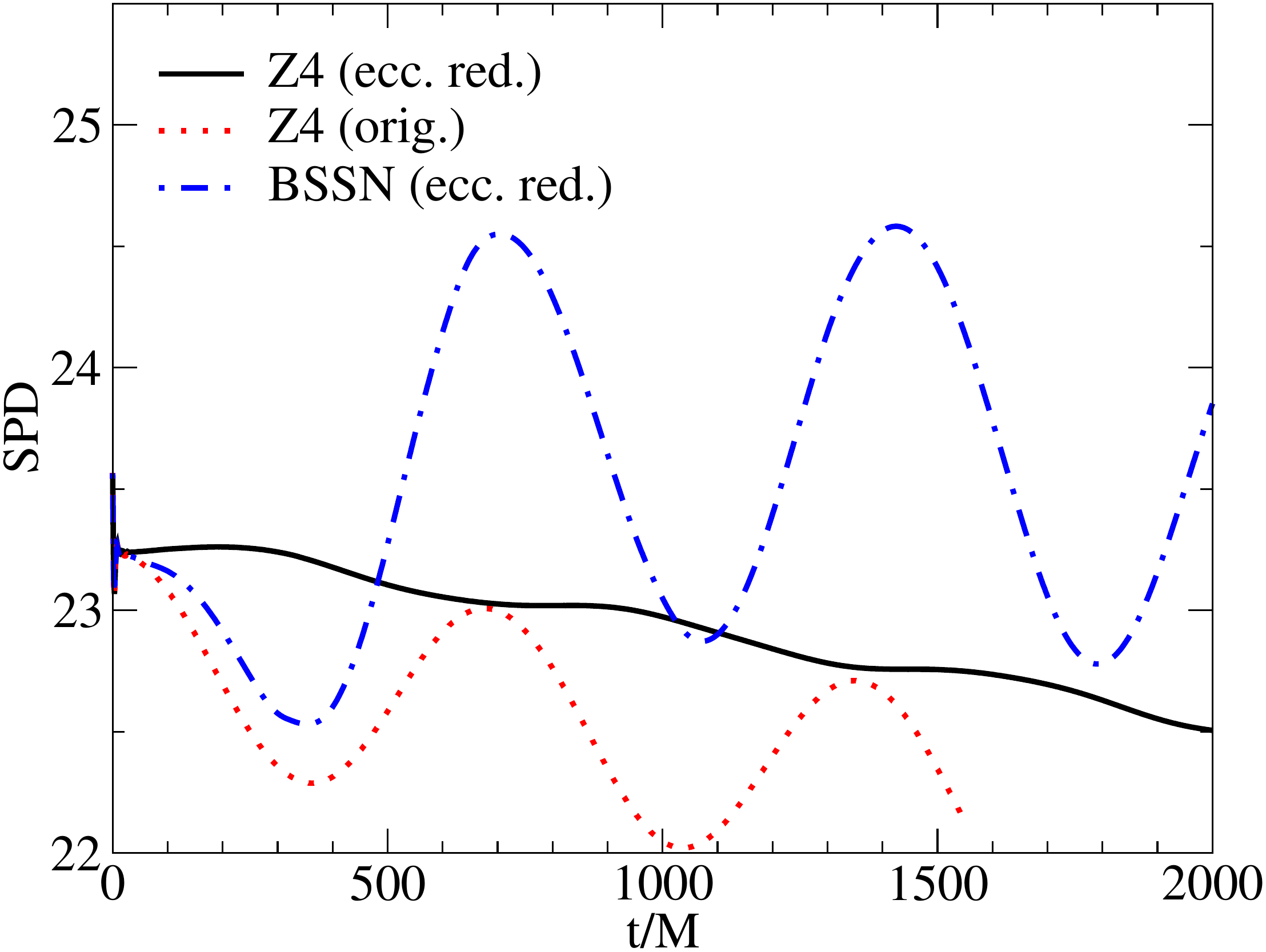}
  \caption{The SPD versus time as calculated using a
    CCZ4 evolution of the new data both before and after the
    eccentricity reduction procedure. For reference, a BSSN evolution
    of the low-eccentricity data is also given. Constraint violation
  drives the eccentricity for this latter case.}
  \label{fig:ecc}
\end{figure}

As shown in Fig.~\ref{fig:geod}, the transition from
analytical to numerical evolutions is relatively smooth, and, as shown above,
evolutions with CCZ4 drive the constraint violations down to
acceptable levels (i.e., within a factor of 10 of the levels obtained by evolving the
constraint satisfying Bowen-York data). The last step required for a successful
continuation of the evolution is to ensure that the binary remains
quasicircular (the PN inspiral used to generate the data is
quasicircular). To accomplish this, we 
applied the eccentricity reduction procedure of Ref.~\cite{Pfeiffer:2007yz} to our data (we
found that we needed to set $M \delta\Omega_{\rm orb} = 7.88515\times10^{-6}$ and
$\delta \dot r =  -1.54103\times 10^{-4}$). After three iterations,
we were left with a residual
eccentricity of $e=0.002$, which was small enough for this test (see,
Fig.~\ref{fig:ecc}).  In Fig.~\ref{fig:ecc}, we show the SPD versus
time for both CCZ4 and BSSN evolutions of the eccentricity-reduced
data, we also show a CCZ4 evolution of the original data. From the
figure, the nonphysical dynamics (overall increase in radius) of the
BSSN evolution is apparent. Note that we implement the eccentricity
reduction by changing the initial orbital inspiral rate $\dot r$ and
orbital frequency $\Omega_{\rm orb}$ used in the PN equations of
motion. Changes to the inner-zone and far-zone metrics are
automatically handled by the matching procedure.

The
eccentricity reduction here is complicated by the fact that the amount of
constraint-violating fields  absorbed by the BHs changes as the
trajectories are modified. This in turn, leads to a more complicated
dependence of the eccentricity on the orbital parameters than is seen
for constraint-satisfying data.

\subsection{Waveform}
\begin{figure}
    \includegraphics[width=.98\columnwidth]{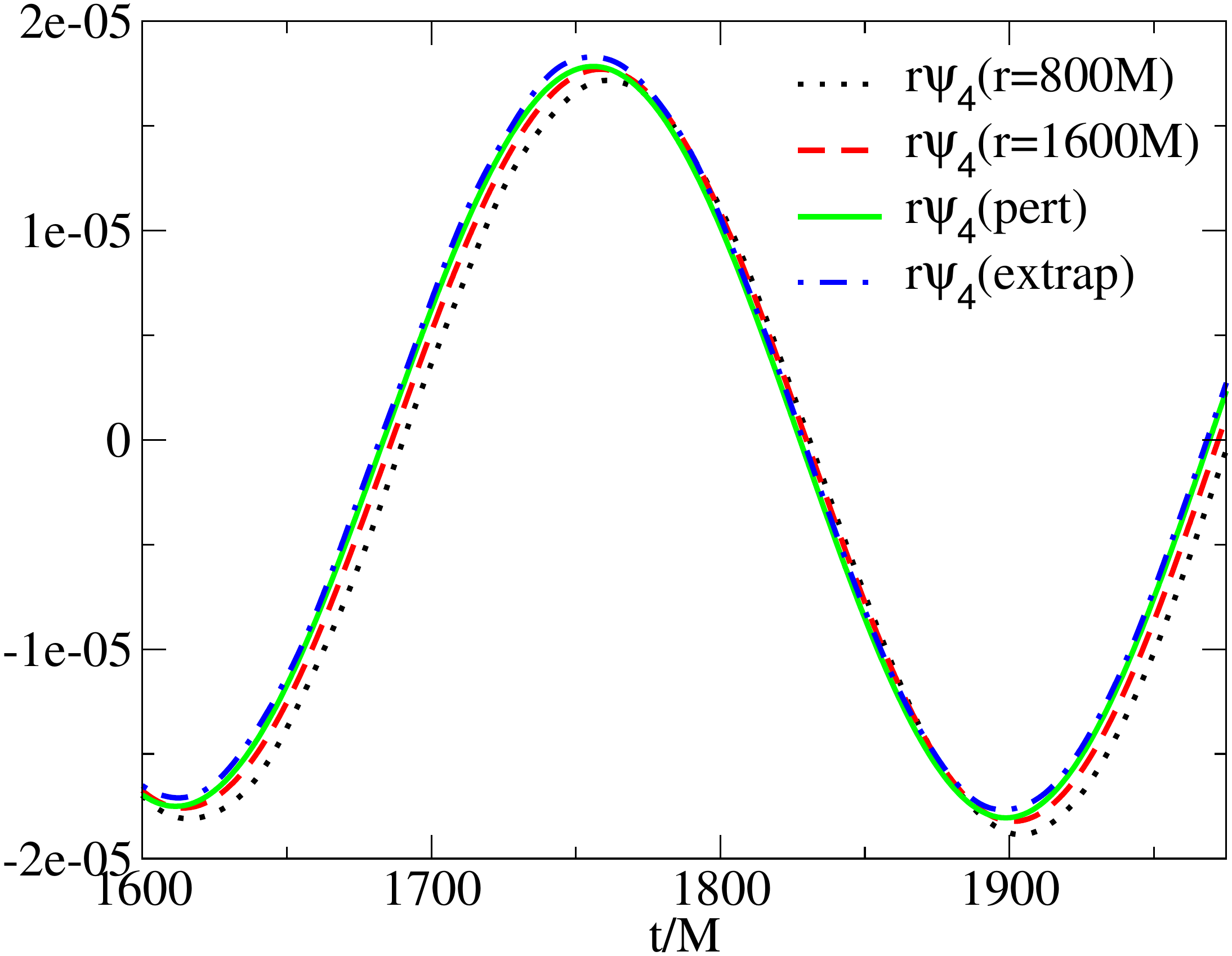}
    \caption{The $(\ell=2,\, m=2)$ mode of the waveform ($r \psi_4$)
      extracted at $r=800M$ and $r=1600M$, a linear extrapolation of
      these to $r=\infty$, and an extrapolation to $r=\infty$ using
      perturbative techniques. Both the standard extrapolation in $r$
    and the perturbative approach give very similar
  waveforms.}\label{fig:extrap}
\end{figure}

\begin{figure}
      \includegraphics[width=.98\columnwidth]{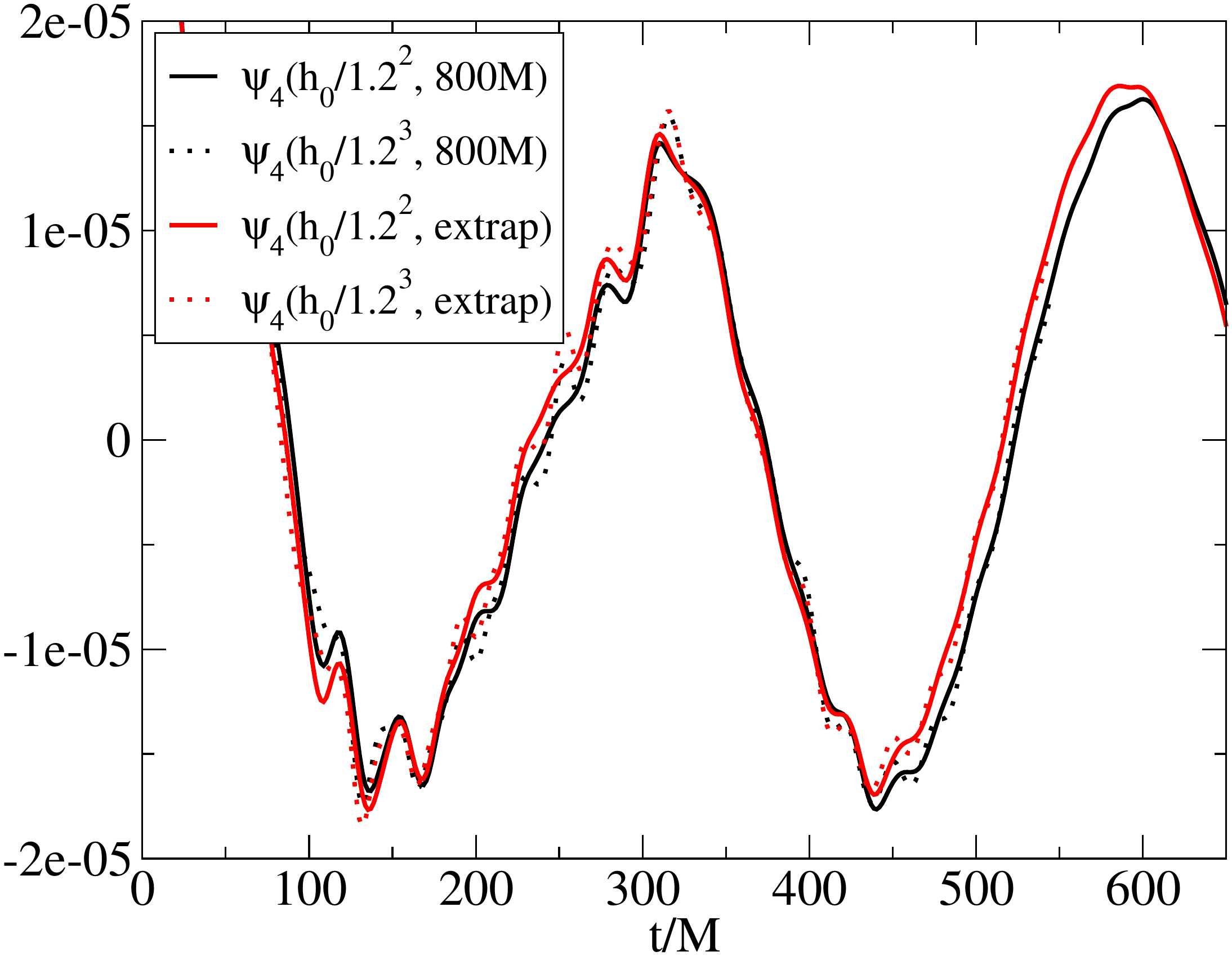}
      \caption{ The $(\ell=2,\, m=2)$ mode of the waveform ($r
        \psi_4$) extracted at $r=800M$ at two resolutions. Both
        the raw waveform and the extrapolation to infinity are shown.
        The dominant error is the extrapolation error which manifests
      itself predominately as a phase error.}\label{fig:wave_accuracy}
\end{figure}

The waveforms presented below are relatively short (due to the expense
of running the simulation to merger, which would take about six months on
80 Opteron cores). We will be comparing the numerical waveforms to PN
waveforms. For these ``short'' runs, the dominant error is due to
finite extraction radius.
 In Fig.~\ref{fig:extrap}, we show a ``late'' segment of the
waveform extracted at $r=800M$, $r=1600M$, and a linear extrapolation (in
$l=1/r$) to $r=\infty$ from these two waveforms, as well as the
extrapolation to $r=\infty$ using the perturbative approach of
Refs.~\cite{Nakano:2015pta, Nakano:2015rda}. As expected, the dominant
errors due to finite radius are phase errors. Finally,
in Fig.~\ref{fig:wave_accuracy}, we show the waveform (post-initial burst)
extracted at $r=800M$ for the resolutions $h=h_0/1.2^2$ and
$h=h_0/1.2^3$. We also show the extrapolation of these waveforms using
the techniques of Refs.~\cite{Nakano:2015pta, Nakano:2015rda}. As can
be seen, the
dominant error in the waveform is the phase error due to finite
extraction radius.

\section{Comparison to PN}\label{sec:cmp_to_pn}

To gauge the accuracy of our transition from analytical to numerical
evolutions, we compare the subsequent dynamics of the binary with the
predictions of PN.

\begin{figure}
\includegraphics[width=.98\columnwidth]{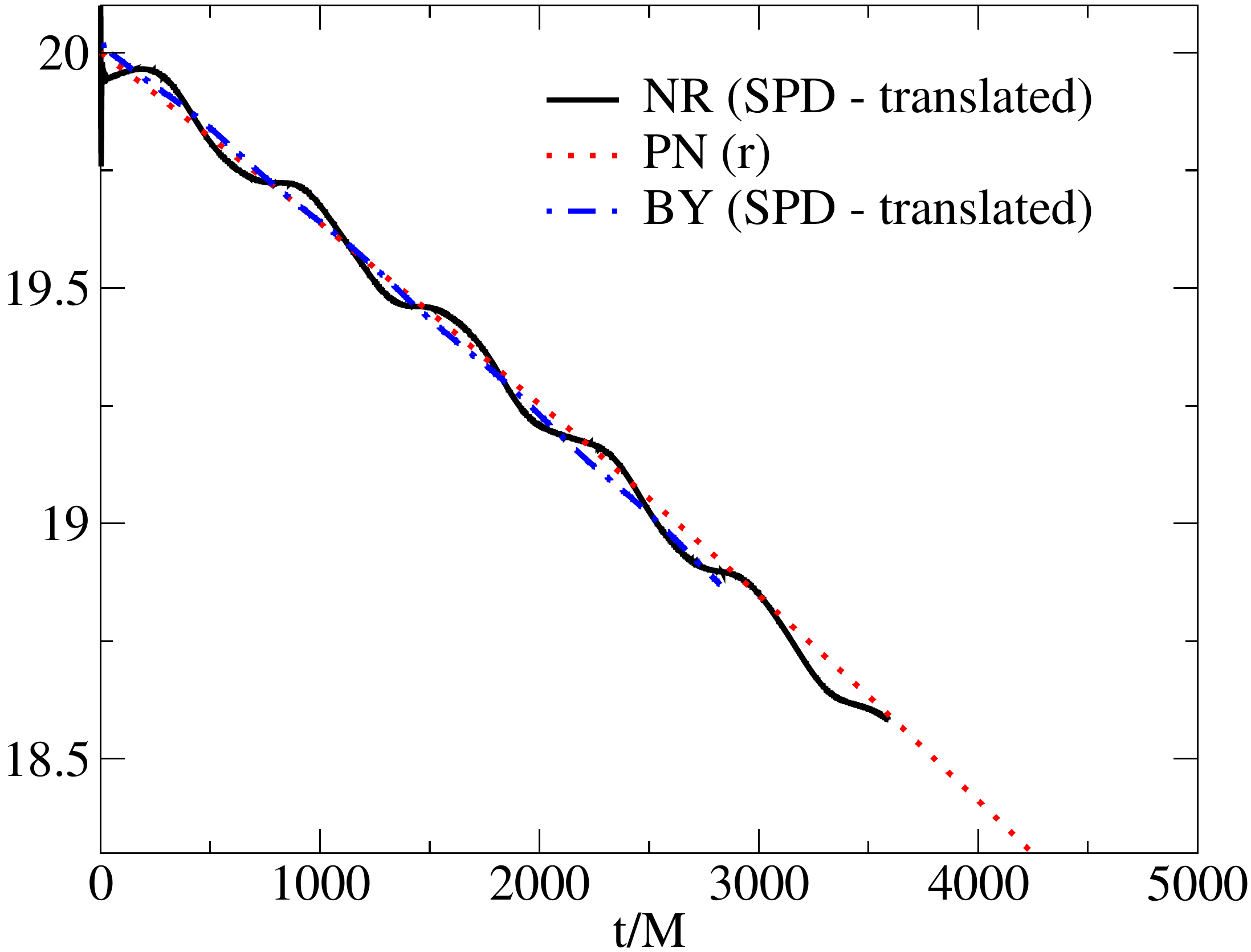}
\caption{The SPD 
and PN separation versus time as calculated using a
CCZ4 evolution of the new data and an older BSSN evolution of
Bowen-York data. The SPD is always larger than the coordinate
(and hence PN) separations. We shift the SPDs downward by $3.25M$ so
that they agree with initial PN separation at $t=0$.
}\label{fig:Z4_PN_sep_cmp} \end{figure}

In Fig.~\ref{fig:Z4_PN_sep_cmp}, we
show the SPD versus time and PN separation versus time. Since the SPD
at $t=0$ is larger than $20M$ we translate the SPD vertically. Note
that the SPD is not expected to be equal to the PN separation. The SPD
includes effects due to the nonflatness of the spatial metric and
measures how distant the two horizons are from each other,
while the PN separation extends from the center of one BH to the
other and the proper separation corresponding to this would not be
finite.
While
it is interesting that the numerical SPD matched the PN separation
reasonably well, 
these are not gauge-invariant quantities. We also
show the SPD for an equivalent Bowen-York simulation (evolved with
BSSN) first reported in Ref.~\cite{Lousto:2013oza}. The Bowen-York
run also had the eccentricity reduction procedure applied.

\begin{figure}
  \includegraphics[width=0.98\columnwidth]{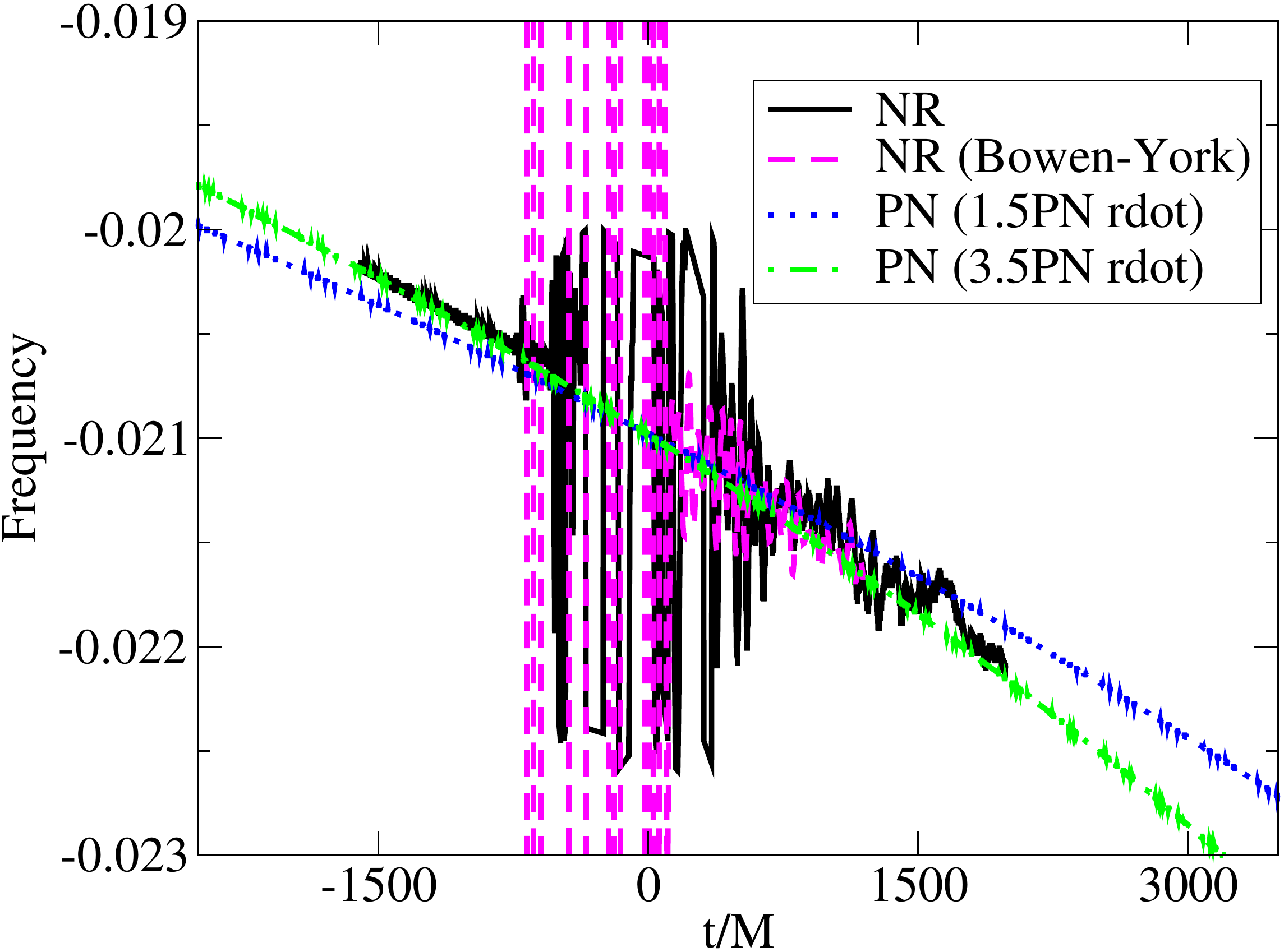}
  \includegraphics[width=0.98\columnwidth]{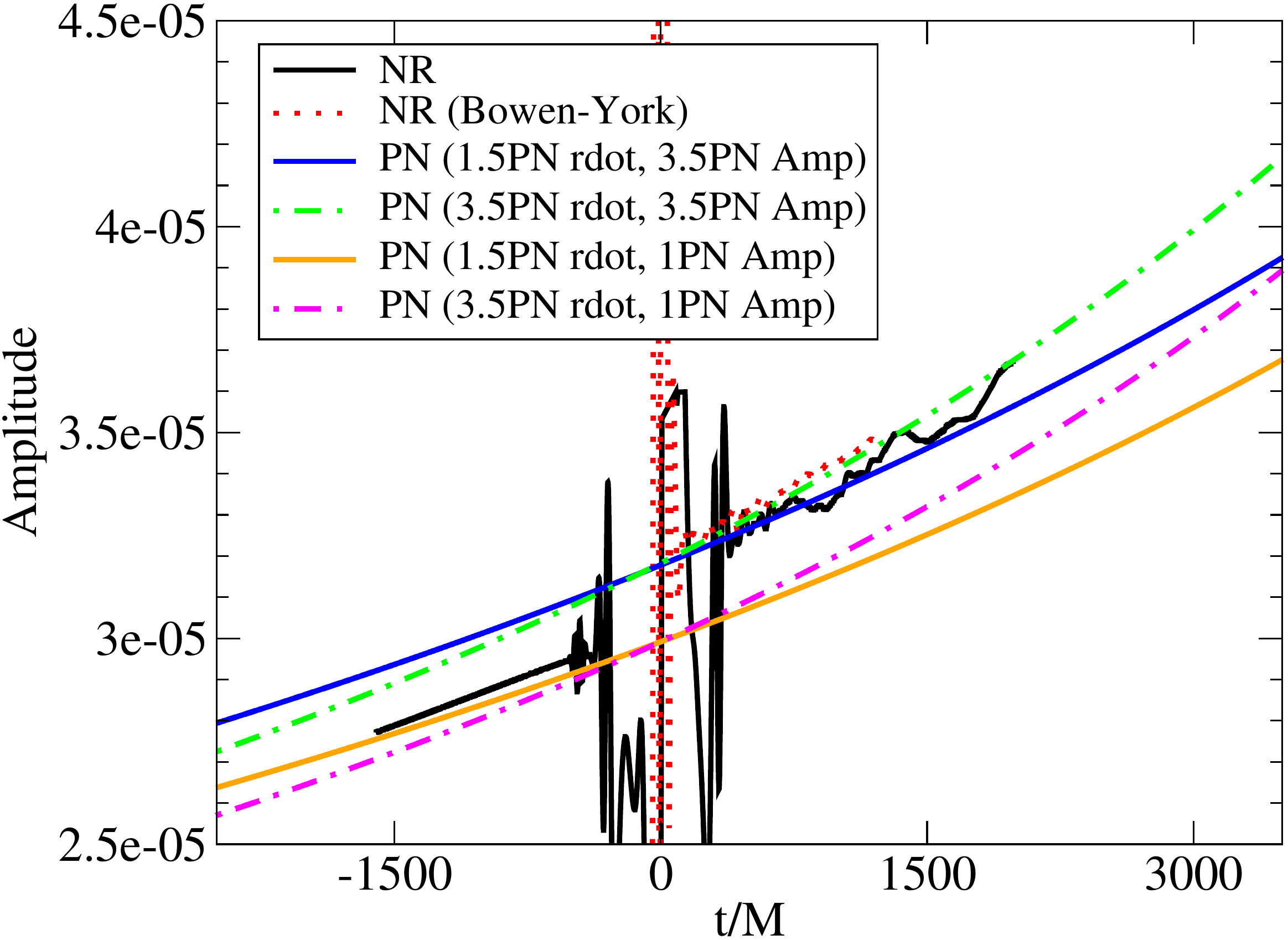}
  \caption{The frequency and magnitude of the $(\ell=2, m=2)$ mode of
$r \psi_4$ as measured at $r=1600M$ in the full numerical simulation
and various PN predictions. Here we use either the 1.5PN or 3.5PN
expressions for $\dot r$ and either the full 3.5PN expression for 
$h_{22}$ (as a function of $r$ and $\omega$) of Faye \etal~\cite{Faye:2012we},
or we truncate to 1PN order. We use the 3PN
expression for $\omega$ in all cases.}
\label{fig:Z4_PN_Wave_cmp}
\end{figure}

To have a more gauge-invariant measure of the accuracy of the
evolution, we compare the waveform (as extracted at $1600M$) with the
3.5PN prediction for quasicircular orbits~\cite{Faye:2012we} (similar to what was done in
Refs.~\cite{Lousto:2013oza} and~\cite{Campanelli:2008nk}). 
  All waveforms are shown in
Fig.~\ref{fig:Z4_PN_Wave_cmp}.

When extracting at $r=1600M$, we get very
good agreement between the {\it raw} $(\ell=2,\, m=2)$ mode of $\psi_4$ 
and the extrapolation to infinity using the techniques
of Ref.~\cite{Nakano:2015pta}. Note that the numerical waveform prior to the burst
of radiation is purely a function of the initial data. The initial
data used the 3.5PN equations of motion; thus the agreement in the
frequency at early times with 3.5PN is expected. 
On the other hand, the PM metric in the FZ contains terms up to 2.5PN order,
which naturally leads to a lower-order approximation for the wave
amplitude, since it depends not only
on the orbital parameters, but also on the metric
perturbation order.
After the initial data burst, the waveform becomes noisier, but the agreement
with 3.5PN is still quite good. The numerical waveform amplitude, however,
seems to be closer to the average of 1.5PN and 3.5PN.

\begin{figure}
  \includegraphics[width=.99\columnwidth]{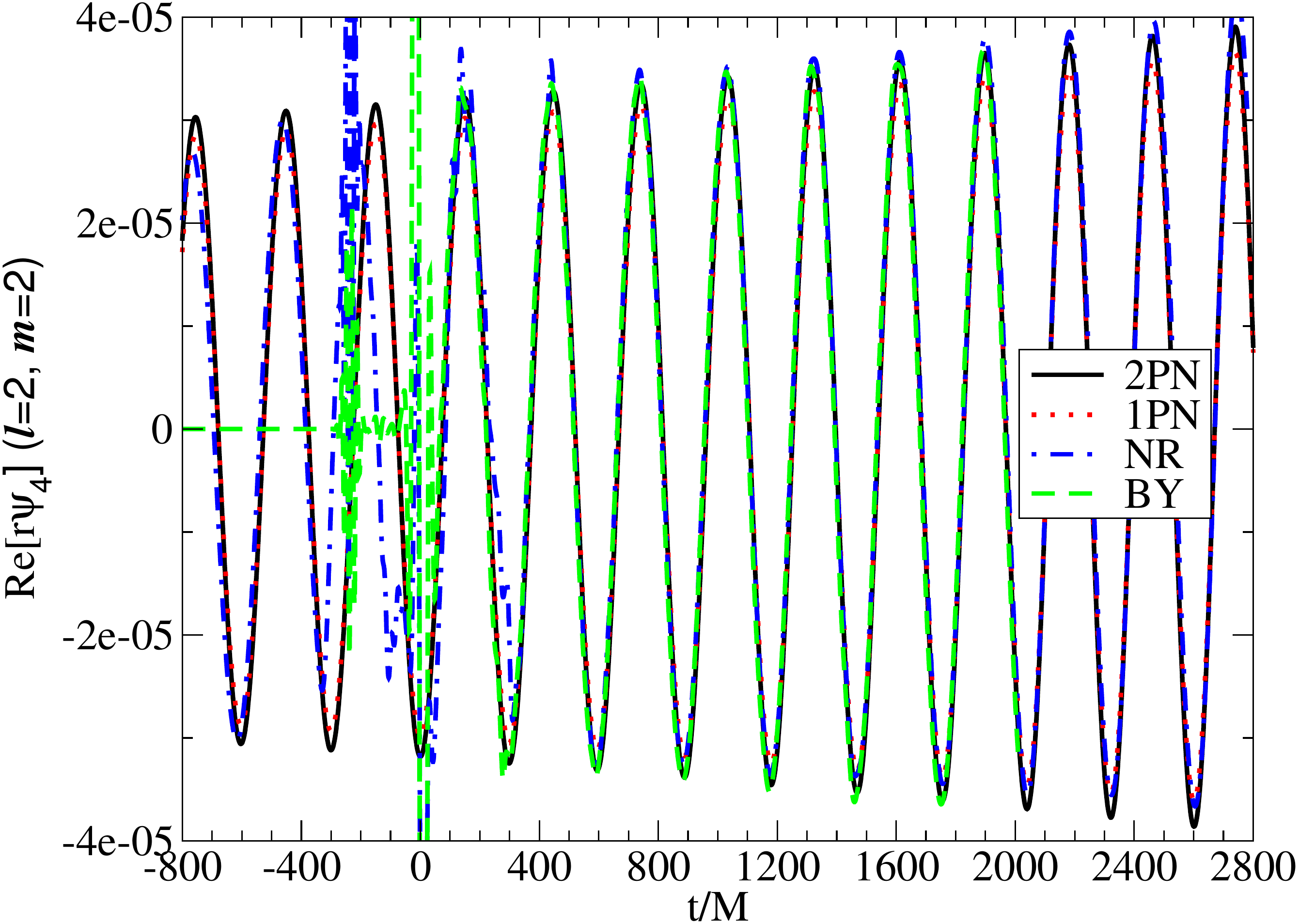}
  \caption{The numerical waveform from the new data, a numerical
    waveform from equivalent Bowen-York data, 
and the PN waveform.
    Here the numerical waveforms
    are shifted by $r^*$ (the tortoise coordinate at the extraction
    radius),
and the PN waveform is unshifted.  The phase agreement is good
after $t=0$ but breaks down
prior to the initial data pulse (for the new data) despite the fact that the
NR frequency is in close agreement with PN for the whole waveform. The jump in
phase between the early and late part of the waveform is likely due to
the use of the flat-space retarded time in constructing the early
waveform.}
\label{fig:dephasing}
\end{figure}

One important note is that the PN waveform given in the initial data
is slightly out of phase with the resulting numerical waveform, as
shown in Fig.~\ref{fig:dephasing}.  That
is to say, after translating the waveform in time by $r^*$ (the
tortoise coordinate of the extraction observer), the PN and NR
waveforms agree quite well for the part of the waveform 
after the initial burst has hit the extraction sphere (in the plot,
this would be from $t=0$ to $t=2400M$).
However, prior to this burst arriving at the observer (in the plot,
prior to $t=0$), the PN and NR
waveforms are out of phase by $0.255$ radians. This initial
part of the NR waveform is produced by the far-zone metric in the
initial data, while the latter part of the waveform is produced by
subsequent fully nonlinear binary dynamics.
This will have repercussions if one wants
to smoothly attach a PN waveform to the numerical waveform. It is
important to note that other than a translation by the tortoise
coordinate $r^*$ corresponding to the extraction radius, the NR and PN
waveforms have not been translated.

The phase error itself can be explained by how we construct the metric
in the far zone. In the far zone, the metric at some point at a
(coordinate) distance $r$ from the origin depends on the dynamics of
the binary at a retarded time given by the light propagation time from
the binary to that point. We use the expression $t_{\rm ret} = t-r$,
which is the flat-space retarded time. A more accurate expression
would include the mass of the spacetime. For a Schwarzschild BH,
in harmonic coordinates, the retarded time would be
\begin{equation}
  t_{\rm ret}^{\rm Sch.} = t-\left[(r+M) +2M
\log\left(\frac{r+M}{2M}-1\right)\right] \,,
\end{equation}
where $M$ is the mass of the spacetime.
We thus find that for a given waveform frequency $\omega$, using the
flat-space retarded time will introduce a phase error of approximately
\begin{equation}
  \delta \phi = \omega \left[M +2M
\log\left(\frac{r+M}{2M}-1\right)\right] \,.
\end{equation}
Since the binary's orbital period here is $M \Omega_{\rm orb} \approx
0.01$, and the $(\ell=2, m=2)$ mode of the waveform has twice this
frequency, we expect a phase error introduced by the flat space
retarded time of $\approx0.287$ radians, which is reasonably close to our
measured phase error of $0.255$ radians.

\begin{figure}
  \includegraphics[width=.98\columnwidth]{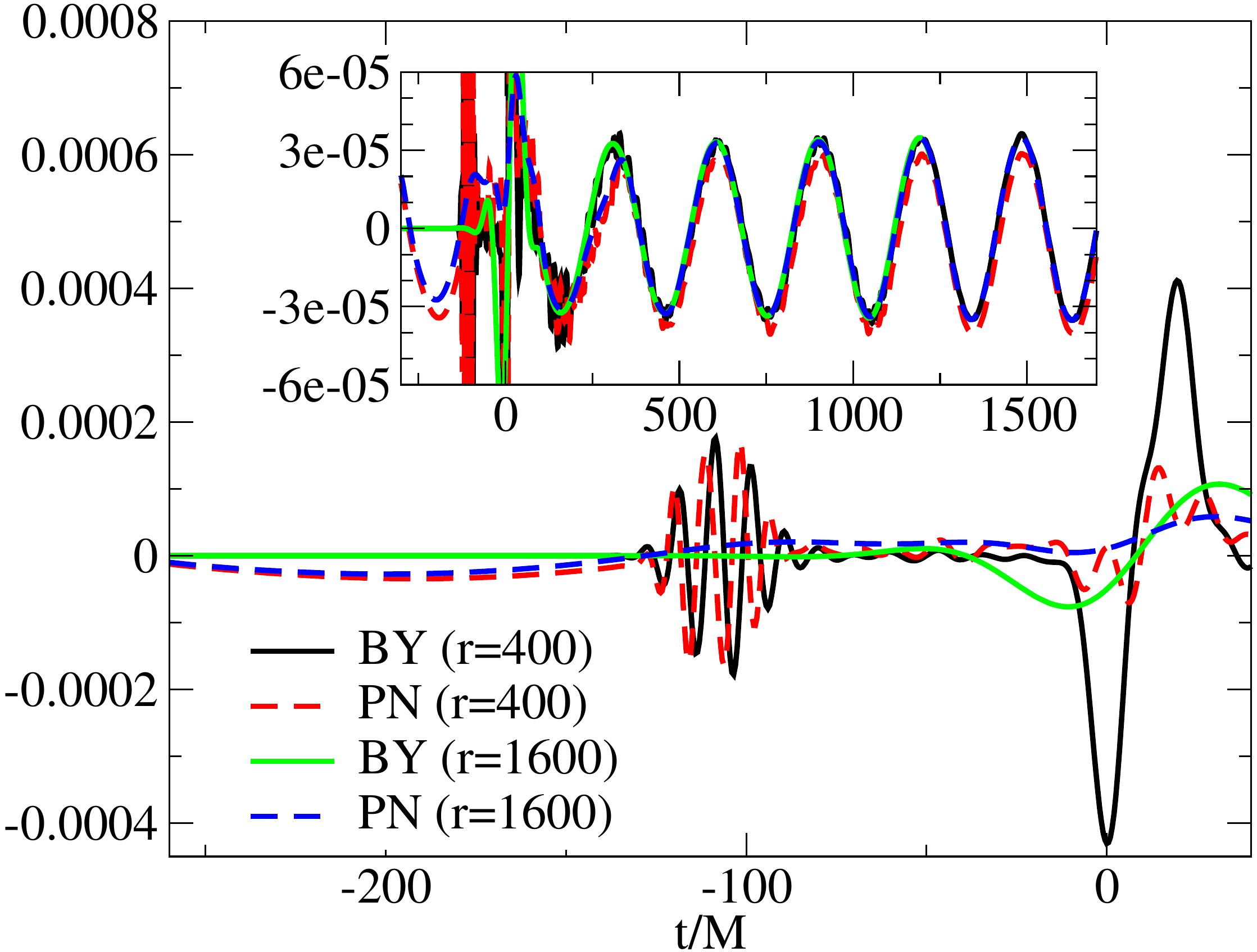}
\caption{A plot of the real part of the $(\ell=2, m=2)$ mode of
$r \psi_4$ (shifted in time by $-r$) for Bowen-York and the
hybrid data here (denoted by PN). The waveforms extracted at $r=400M$
show a high-frequency pulse near $t=-100M$ due to an unresolved gauge
wave. This high-frequency pulse is dissipated away and not visible in
the waveform extracted at $r=1600M$. }\label{fig:pulse}
\end{figure}

One final note concerns the amplitude of the initial data pulse in the
waveform. As seen in Fig.~\ref{fig:pulse}, the initial pulse of
radiation is suppressed relative to equivalent Bowen-York data. At
$r=1600M$, the suppression is roughly a factor of $2$, while at the
$r=400M$ extraction radius, the suppression is closer to a factor of
3. 
This is mostly due to the fact that we have initial data which model the 
astrophysical BHB system better and therefore possess less spurious 
radiation content when compared to the conformally flat BY initial data.
Here, because of the resolution of the grid at $r=1600M$, the
high-frequency gauge pulse (near $t=-500M$) is completely dissipated
away. The high-frequency components of the initial data pulse are
similarly suppressed.

\section{Discussion}\label{sec:discussion}

In order to perform accurate GRMHD simulations of gas accreting onto a
BHB, including the minidisks around each BH, we need a
spacetime that is accurate for the entire lifetime of the
binary, i.e., from the slow inspiral at extremely large separation all
the way to merger and relaxation of the remnant BH. 
Our goal, therefore, is to produce a
four-dimensional metric that is
accurate outside the horizons at all times, of sufficient
smoothness that timelike geodesics vary smoothly (i.e., are
$C^\infty$) except at a single transition time where they are $C^2$,
and the subsequent binary dynamics should match PN predictions to a
high 
degree in the vicinity of this transition time.
To do this, we
extended an analytic metric that is accurate when the
binary's separation is $D\gg M$ by continuing the evolution
using fully nonlinear numerical
techniques for closer separations.

The main questions we addressed here
concerned how we can accurately transition from using an
analytically evolved spacetime metric to a
fully nonlinear numerically evolved metric that describes the binary
during the rapid plunge and merger. At the transition, we used the
analytical spacetime metric to construct initial data for a subsequent
numerical evolution (as was previously done in
Reifenberger and Tichy~\cite{Reifenberger:2012yg}). Our work builds
upon Reifenberger and Tichy in two main ways. We start from an
analytic spacetime that can be extended arbitrarily far into the past,
and we can thus compare dynamics of particles pre- and post-transition.
We also perform the transition at a binary
separation of $D\sim 20M$, where the binary's dynamics are still well
described by PN theory and errors introduced in the gas dynamics by the approximate metric
are {\it washed} out by MHD turbulence (see Ref.~\cite{Zilhao:2014ida} for an
analysis of MHD evolutions on this analytical background for
various separations).

In order for the transition from an analytical
evolution to a numerical one to be smooth enough, the binary's orbital dynamics could not
change significantly as a result of the transition.  The binary's
dynamics in the fully nonlinear numerical  simulation had two main
sources of error. First, constraint violations led to rapid
unphysical oscillations in the orbital separation. Second, small
errors in the PN expressions for the orbital angular
momentum and inspiral rate led to eccentricity in the binary.  We
were able to ameliorate the first source of error
by 
evolving with the constraint-damping CCZ4~\cite{Alic:2011gg}
formulation of the Einstein
equations, which causes constraint violations to rapidly propagate
away from the BHs, significantly reducing unphysical binary dynamics.
In addition, by adding small changes to
the initial inspiral rate and orbital frequency, we significantly
reduced the eccentricity of the numerical binary using the techniques
of Ref.~\cite{Pfeiffer:2007yz}.

We subsequently found that the NR evolution leads to the expected
gravitational waveform, orbital frequency, and binary inspiral rate (to
within the truncation error of the simulation). The remaining error
we found is a phase error in the early part of the waveform. This
phase error is about $0.255$ radians. We ascribe this error to our use of
the flat-space retarded time in the far-zone. By not including effects
due to the mass of the spacetime  we generate phase errors of the
order of $0.287$ radians in the waveform. This error can itself be
ameliorated by using the Schwarzschild retarded time when constructing
the far-zone metric, which is something we will explore in an upcoming
paper.

\begin{acknowledgments}

We thank Carlos Lousto for a careful reading of this
manuscript. We thank Carlos Lousto, Zachariah Etienne,
Nicol\'{a}s Yunes, and Ian Hinder for
helpful discussions.  The authors are supported by NSF
Grants   No. AST- 1516150, No. ACI-1516125, No. PHY-1305730, No.
PHY-1212426, No. PHY-1229173, No. AST-1028087,
No. OCI-0725070 (PRAC subcontract  2077-01077-26), No. OCI-0832606.
Computational
resources were provided by XSEDE allocation No. TG-PHY060027N, and by
NewHorizons and BlueSky Clusters at Rochester Institute of Technology,
which were supported by NSF grant No. PHY-0722703, No. DMS-0820923,
No. AST-1028087, and No. PHY-1229173. B.~C.~M.~ is supported by the LOEWE-Program 
in HIC for FAIR.
H.~N. acknowledges support by 
MEXT Grant-in-Aid for Scientific Research
on Innovative Areas,
``New Developments in Astrophysics Through Multi-Messenger Observations
of Gravitational Wave Sources'', No.~24103006.
M.~Z.\ is supported by grants 2014-SGR-1474, MEC FPA2010-20807-C02-01, MEC FPA2010-20807-C02-02, CPAN CSD2007-00042 Consolider-Ingenio 2010, and ERC Starting Grant HoloLHC-306605. 
\end{acknowledgments}

\bibliographystyle{apsrev4-1}
\bibliography{references}
\end{document}